\documentclass[journal]{IEEEtran}

\usepackage{amsmath,amssymb,amsfonts}
\usepackage{bm}

\usepackage{graphicx}
\usepackage{xcolor}
\colorlet{blue}{black}
\usepackage{array}
\usepackage{booktabs}
\usepackage[caption=false,font=normalsize,
labelfont=sf,textfont=sf]{subfig}

\usepackage{algorithm}
\usepackage{algpseudocode}

\usepackage{textcomp}
\usepackage{stfloats}
\usepackage{placeins}
\usepackage{url}
\usepackage{verbatim,flushend}
\def\BibTeX{{\rm B\kern-.05em{\sc i\kern-.025em b}\kern-.08em
    T\kern-.1667em\lower.7ex\hbox{E}\kern-.125emX}}
\usepackage{balance}
\begin{document}
%\title{Multiuser Channel Estimation for Flexible Intelligent Metasurfaces via Structured Block Term Decomposition}
\title{Multiuser Channel Estimation in FIM Systems:\\ A Structured Block Term Decomposition Approach}

\author{Xi Han\textsuperscript{1,2}, Junqiao Chen\textsuperscript{1}, André L. F. de Almeida\textsuperscript{3}, George C. Alexandropoulos\textsuperscript{4}
\thanks{\textsuperscript{1}X. Han and J. Chen are with the Beijing Key Laboratory on Integration and Analysis of Large-scale Stream Data, School of Information Science and Technology, North China University of Technology, Beijing 100144, China (e-mail:jinlucky333@gmail.com; cjq1251726586@gmail.com). \textsuperscript{2}X. Han is also with the State Key Laboratory of Intelligent Optimized Manufacturing in Mining \& Metallurgy Process. \textsuperscript{3}André L. F. de Almeida is with the Laboratory of Applied Signal Processing and the Wireless Telecom Research Group, Department of Teleinformatics Engineering, Federal University of Ceará, Fortaleza, Brazil (e-mail: andre@gtel.ufc.br). \textsuperscript{4}G. C. Alexandropoulos is with the Department of Informatics and Telecommunications, National and Kapodistrian University of Athens, 16122 Athens, Greece (e-mail: alexandg@di.uoa.gr). Corresponding author: André L. F. de Almeida.}%
\thanks{This work was supported in part by the National Key Research and Development Program of China under Grant 2024YFE0200103, the China Scholarship Council (Grant No. 202508110189), the NSFC under Grant 62001008, and the Beijing Municipal Natural Science Foundation under Grants 4212002 and L192034. This work is partially supported by the National Institute of Science and Technology (INCT-Signals) sponsored by Brazil's National Council for Scientific and Technological Development (CNPq) (406517/2022-3), and FUNCAP (INCT-25255-82587.32.41/64). The research of André L. F. de Almeida is partially supported by CNPq (303356/2025-1).}
}

%\markboth{Journal of \LaTeX\ Class Files,~Vol.~18, No.~9, September~2020}%{How to Use the IEEEtran \LaTeX \ Templates}

\maketitle

\begin{abstract}
\color{blue}
Flexible intelligent metasurfaces (FIMs) create geometry-dependent array responses that provide spatial diversity but complicate multiuser channel estimation. This paper studies uplink estimation using pilot observations collected under multiple prescribed and calibrated morphing configurations. Each user contributes a matrix--vector block that couples its diagonal reference-array response with its morphing-dependent phases, while its third-mode vector is the known pilot scaled by the complex channel gain. We develop a pilot-constrained structured block term decomposition (BTD) estimator whose alternating least-squares updates exploit both the known pilot directions and the diagonal spatial factors. The undeformed configuration sets the columnwise scaling convention, and a unit first-entry normalization assigns the remaining block scale to the channel gain. Matching the recovered factors to the calibrated spatial--morphing manifold then yields the physical angles and configuration-dependent channels. We characterize the rank conditions for unique conditional least-squares updates, distinguish these conditions from global factor and physical-angle identifiability, and summarize the training, coherence-time, and computational requirements. Simulations show accurate two-dimensional angle and channel recovery in the tested settings, equal total pilot-symbol overhead in rigid-panel comparisons, improved steering-vector estimates as morphing diversity increases, and degradation as user loading or pilot correlation increases.
\end{abstract}


\section{Introduction}
\label{sec:introduction}

%\IEEEPARstart{R}{econfigurable} 
Reconfigurable intelligent surfaces (RISs) have emerged as a promising 
technology for realizing programmable wireless propagation environments
by introducing controllable electromagnetic responses into the radio
channel~\cite{AlexandropoulosRIS}. Most
conventional RIS architectures achieve such reconfigurability by tuning
element-wise responses while maintaining
a fixed physical aperture~\cite{10596064}. More recently, beyond-diagonal RIS
architectures have further extended the electromagnetic degrees of
freedom by introducing tunable interconnections among surface elements
\cite{li2023bdris}. Despite these advances, the physical coordinates of
the radiating or scattering elements remain essentially fixed, and the
resulting spatial manifold is therefore determined by a prescribed
array geometry.

Flexible intelligent metasurfaces (FIMs) provide a fundamentally
different form of reconfigurability by introducing controllable
geometric degrees of freedom. Specifically, an FIM allows its elements
to move along the direction normal to the nominal surface, thereby
morphing the three-dimensional aperture and directly modifying its
spatial response. Recent studies have demonstrated the potential of
such geometry reconfiguration for reducing the transmit power in
multiuser systems and enhancing multiple-input multiple-output (MIMO) channel capacity
\cite{an2025fimmiso,an2025fimmimo}. Closely related movable-antenna
architectures have also shown that position-dependent channel
measurements can be exploited to reconstruct spatially varying channels
from a limited number of sampled antenna locations
\cite{xiao2024movable,zhang2024tensor}. These results indicate that
geometric reconfiguration provides a new source of spatial diversity
that cannot be reproduced through conventional response control.

The additional geometric degrees of freedom, however, substantially
complicate channel acquisition. Unlike a rigid array whose steering
response remains fixed during training, an FIM exhibits a
configuration-dependent array manifold since each prescribed
morphing profile changes the physical element coordinates. Estimating
the channel independently for every possible surface shape would
therefore incur excessive training overhead and would fail to exploit
the deterministic coupling among different morphing configurations.
Recent studies have begun to address this issue. A sparsity-based
channel estimator combined with FIM shape optimization was developed
in \cite{jiang2026enhanced}, while interpolation, kernel-based
reconstruction, sparse parametric recovery, and neural operators were
investigated for learning the channel over a continuous deformation
space in \cite{xiao2026fimchannel}. A tensor-based channel estimation
framework was proposed for point-to-point MIMO systems equipped
with morphing FIM arrays, where the multidimensional observations are
modeled through a PARAFAC decomposition and processed by alternating
least squares \cite{romano2026fim}. These works demonstrate the
feasibility of exploiting geometric structure for FIM channel
acquisition, but the problem of separating multiple users having
distinct spatial directions under a common set of calibrated morphing
configurations remains largely unexplored.

More broadly, channel estimation for RIS-assisted communications has
progressively evolved from unstructured least-squares estimation toward
model-aware approaches that exploit sparsity, parametric structure, and
hardware constraints. The fundamental training and identifiability
issues associated with RIS channel acquisition were systematically
reviewed in \cite{shamasundar2023channel}. For sparse propagation
environments, compressed-sensing techniques can substantially reduce
the number of required observations, although their performance may be
sensitive to angular grid mismatch \cite{daryanavardan2025sparse}.
Data-driven super-resolution methods have also been proposed for
recovering high-dimensional RIS channels from reduced measurements
\cite{shen2023deep}. In addition, the emergence of beyond-diagonal RISs
has motivated channel-estimation and training designs that explicitly
account for non-diagonal scattering structures
\cite{li2024bdrisce}. Nevertheless, these methods predominantly assume
fixed physical element coordinates and therefore do not explicitly
capture the deterministic variation of the spatial manifold across the
multiple possible FIM morphing states.

Tensor methods are particularly attractive for this problem since
they preserve the multilinear coupling among spatial dimensions,
training symbols, reconfiguration states, and propagation components.
Beyond RIS-assisted systems, nested tensor models have
also been exploited for semi-blind joint channel and symbol estimation
in UAV-assisted MIMO systems \cite{han2021semiblind} and for information
monitoring through tensor-contraction-based receivers \cite{han2022monitoring}.
For conventional RIS-assisted MIMO systems, PARAFAC modeling has been
used to decouple the constituent channels from structured pilot
observations \cite{dearaujo2021tensor,9366805,9814591}, and related tensor formulations
have enabled semi-blind joint channel and data estimation
\cite{dearaujo2023semiblind,9786794} as well as robust channel recovery under
RIS hardware imperfections \cite{gomes2023imperfections}. Coupled
tensor decompositions have subsequently been developed for double-RIS
systems \cite{nwalozie2025double}. Tensor-based joint channel and symbol
estimation with subspace-based parameter extraction has further been
investigated for multi-RIS uplink MIMO systems \cite{han2026multiris}.
The channel estimation problem
for beyond-diagonal RISs has recently been formulated as a block Tucker
decomposition that explicitly exploits the underlying block-Kronecker
structure \cite{dealmeida2025bdris}. Tensor-based parametric methods
have also been applied to low-rank mmWave channels
\cite{ardah2021trice,zheng2022compressed}, structured THz channels
\cite{asim2025thz}, multidimensional angular-parameter estimation
\cite{asim2025twod}, and joint channel estimation and activity
detection in RIS-assisted massive connectivity
\cite{cao2025massive}. Collectively, these studies demonstrate that
substantial estimation gains can be achieved when the selected tensor
model is matched to the physical structure of the underlying wireless
system.

{\color{blue}The multiuser FIM model considered here differs from the tensor models summarized above. Each angularly separated user has its own reference-array steering vector and morphing-dependent response, so the observation cannot generally be represented by a single shared spatial factor. Instead, each user contributes a matrix--vector block: its matrix factor jointly describes its spatial and morphing responses, and its vector factor is  its known pilot sequence. This strmotivates a block term decomposition (BTD), unused which represents a tensor as a sum of components with prescribed multilinear ranks \cite{delathauwer2008btd}. A generic BTD, however, does not by itself exploit several forms of prior information available here: the pilot directions are known, each spatial factor is diagonal, and the undeformed configuration provides a physical reference for fixing the factor scaling. Moreover, algebraic factor recovery alone does not ensure reliable angle estimation, which also depends on the calibrated FIM manifold and on the conditioning of the joint pilot--spatial--morphing signatures. The methodological gap addressed in this paper is therefore the joint exploitation of user-specific block structure, known pilots, diagonal spatial factors, and We organize the multi-configuration observationsuser FIM channel estimation.}

{\color{blue}To address e develop a pilot-constrained structured FIM-BTD framework for uplink channel estimation. We organize the multi-configuration observations as a third-order tensor and model each user's contribution as a structured block coupling its spatial response, morphing response, complex gain, and known pilot. The estimator preserves the pilot directions and diagonal spatial structure during factor recovery, then uses the calibrated FIM manifold to estimate the physical angles and reconstruct the channels. The main contributions are as follows.}

\begin{itemize}

\item {\color{blue}We formulate the multi-configuration uplink observations as a user-specific structured BTD in which each block couples a diagonal reference-array factor, a morphing factor, and a known pilot direction. This structure incorporates FIM measurement physics directly into the decomposition.}

\item {\color{blue}We develop alternating structured least-squares updates that retain the known pilot directions and diagonal spatial factors. We use the undeformed configuration and a unit first-entry convention to fix factor scaling, then match the recovered factors to the calibrated spatial--morphing manifold to estimate angles and reconstruct configuration-dependent channels.}

\item {\color{blue}We characterize the rank conditions for uniqueness of each conditional least-squares update and distinguish them from global factor uniqueness and physical-angle identifiability. We also discuss pilot conditioning, morphing diversity, coherence-time requirements, and computational and memory costs.}

\item {\color{blue}Simulations compare the proposed estimator with least-squares baselines and demonstrate two-dimensional angle estimation and channel reconstruction in the tested user settings. The rigid-panel comparison uses the same total pilot-symbol overhead. The results also show how estimation accuracy varies with morphing diversity and user loading, and how highly correlated pilots degrade performance through ill-conditioning.}

\end{itemize}

\section{System Model and Tensor Observation}
\label{sec:system_model}

\subsection{FIM Geometry and Uplink Training Model}
\label{subsec:fim_geometry}

\begin{figure*}[t]
	\centering
	\includegraphics[width=0.98\textwidth]
	{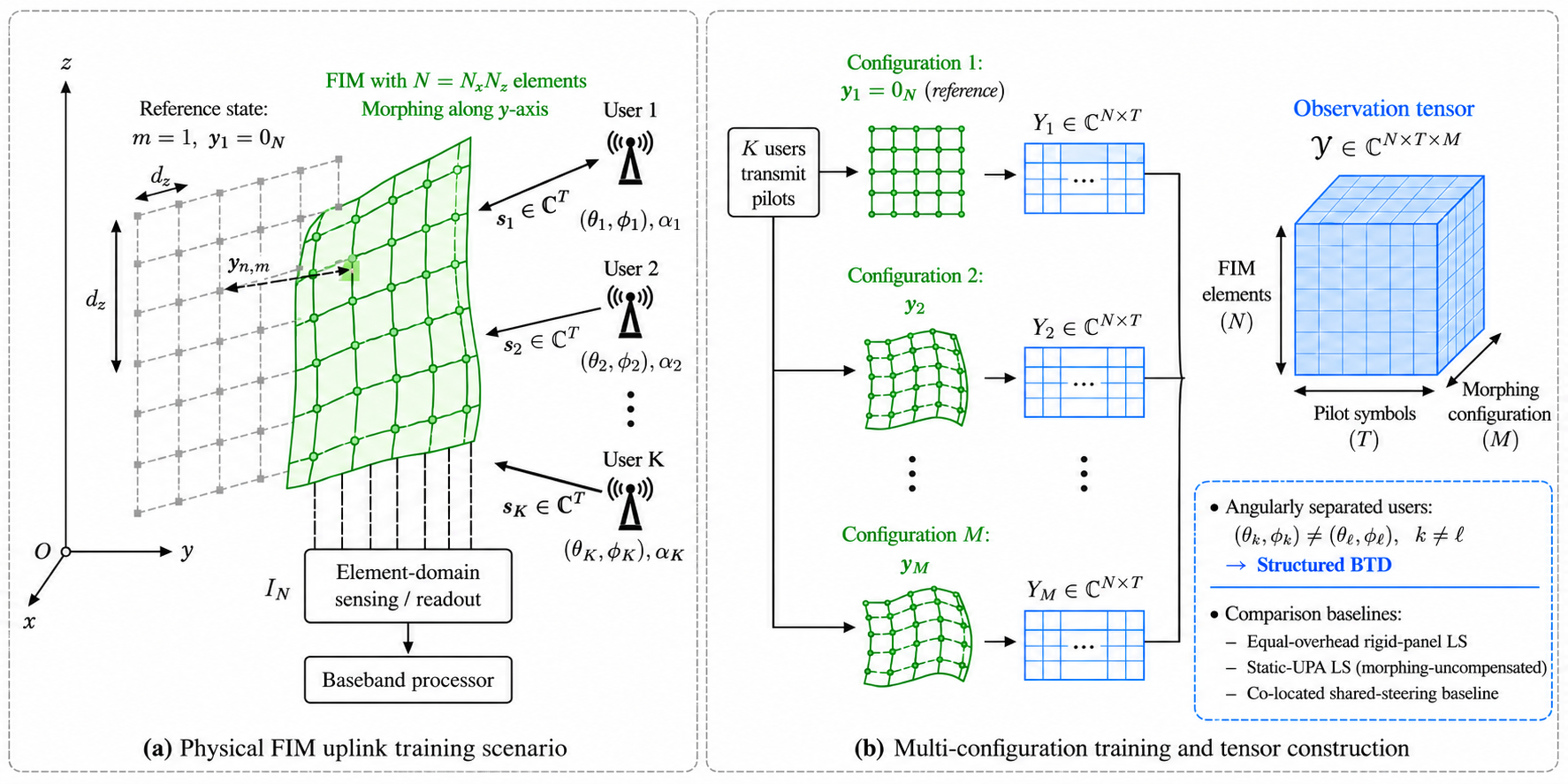}
	\caption{Schematic illustration of the considered multiuser uplink
		training framework with a sensing-enabled flexible intelligent
		metasurface (FIM). 
		(a) Physical FIM uplink training scenario, where the FIM elements
		undergo prescribed displacements along the $y$-axis and the
		angularly separated users transmit known pilot sequences.
		(b) Multi-configuration training and tensor construction, where the
		received pilot matrices collected under $M$ calibrated morphing
		configurations are stacked into the third-order observation tensor
		$\mathcal{Y}\in\mathbb{C}^{N\times T\times M}$ for structured
		FIM-BTD processing.}
	\label{fig:FIM_system_model}
\end{figure*}

We consider the narrowband, block-fading uplink shown in Fig.~\ref{fig:FIM_system_model}, comprising a sensing-enabled FIM and $K$ single-antenna users. The users transmit known pilot sequences, and calibrated sensing and readout chains acquire the signal at each FIM element. We model each user's channel with one dominant far-field path, parameterized by a complex gain and an angular pair.

% ============================================================
% System model figure
% ============================================================

The FIM contains $N=N_xN_z$ electromagnetic elements, where
$N_x$ and $N_z$ denote the numbers of elements along the $x$- and
$z$-axes, respectively. Each element is assumed to be connected to a
calibrated sensing and readout chain, such that its complex baseband
signal can be individually acquired. Hence, the receiver has direct
access to the $N$-dimensional element-domain observation, or
equivalently, the element-domain readout matrix is assumed to be
$\mathbf I_N$. Architectures employing analog combining or compressed
element-domain readout are beyond the scope of this work.

In the undeformed reference configuration, the FIM lies on the
$xz$-plane. The element located at the $p$th position along the
$x$-axis and the $q$th position along the $z$-axis is indexed as
\begin{equation}
n=(p-1)N_z+q,
\qquad
p=1,\ldots,N_x,\quad q=1,\ldots,N_z .
\label{eq:element_index}
\end{equation}
Let $d_x$ and $d_z$ denote the inter-element spacings along the two
axes. The corresponding reference coordinates are
\begin{equation}
x_p=(p-1)d_x,
\qquad
z_q=(q-1)d_z.
\label{eq:reference_coordinates}
\end{equation}

During channel training, the FIM sequentially employs $M$ prescribed
and calibrated morphing configurations. In the $m$th configuration,
the elements retain their reference coordinates along the $x$- and
$z$-axes and are displaced only along the $y$-axis. Hence, the
position of the $n$th element is
\begin{equation}
\mathbf r_{n,m}
=
[x_p,\;y_{n,m},\;z_q]^T,
\qquad m=1,\ldots,M,
\label{eq:element_position}
\end{equation}
where $y_{n,m}$ denotes the out-of-plane displacement of the $n$th
element. The displacement profile of the $m$th morphing configuration
is collected as
\begin{equation}
\mathbf y_m
=
[y_{1,m},\ldots,y_{N,m}]^T
\in\mathbb R^N .
\label{eq:morphing_profile}
\end{equation}
The morphing profiles $\{\mathbf y_m\}_{m=1}^{M}$ are assumed to be
known at the receiver. Deformation-dependent variations in the
element radiation patterns, local orientations, polarization responses,
mutual coupling, and path amplitudes are assumed to be negligible or
compensated through calibration.

Consider a far-field plane wave characterized by elevation angle
$\theta$ and azimuth angle $\phi$. The elevation angle is measured
from the positive $z$-axis, whereas the azimuth angle is measured
from the positive $x$-axis in the $xy$-plane. The corresponding unit
direction vector is
\begin{equation}
\mathbf u(\theta,\phi)
=
[\sin\theta\cos\phi,\;
\sin\theta\sin\phi,\;
\cos\theta]^T .
\label{eq:direction_vector}
\end{equation}
Let $\lambda$ denote the carrier wavelength and
$\kappa=2\pi/\lambda$ the wavenumber. Using a fixed receive phase
convention, the response of the $n$th FIM element under the $m$th
morphing configuration is
\begin{align}
[\mathbf a_m(\theta,\phi)]_n
&=
e^{j\kappa\mathbf r_{n,m}^T\mathbf u(\theta,\phi)}
\nonumber\\
&=
e^{j\kappa(x_p\sin\theta\cos\phi+z_q\cos\theta)}
e^{j\kappa y_{n,m}\sin\theta\sin\phi}.
\label{eq:fim_element_response}
\end{align}
The reference-array responses along the $x$- and $z$-axes are defined
as
\begin{equation}
[\mathbf a_x(\theta,\phi)]_p
=
e^{j\kappa(p-1)d_x\sin\theta\cos\phi}
\label{eq:x_steering}
\end{equation}
and
\begin{equation}
[\mathbf a_z(\theta)]_q
=
e^{j\kappa(q-1)d_z\cos\theta},
\label{eq:z_steering}
\end{equation}
respectively. The steering vector of the undeformed uniform planar
array (UPA) is therefore
\begin{equation}
\mathbf a_{\rm UPA}(\theta,\phi)
=
\mathbf a_x(\theta,\phi)\otimes\mathbf a_z(\theta)
\in\mathbb C^N ,
\label{eq:upa_steering}
\end{equation}
where $\otimes$ denotes the Kronecker product.
The phase variations induced by the $m$th morphing profile are
collected in
\begin{equation}
\mathbf e_m(\theta,\phi)
=
e^{j\kappa\sin\theta\sin\phi\,\mathbf y_m}
\in\mathbb C^N ,
\label{eq:morphing_response}
\end{equation}
where the exponential is applied elementwise. Accordingly, the
complete FIM steering vector is expressed as
\begin{align}
\mathbf a_m(\theta,\phi)
&=
\mathbf a_{\rm UPA}(\theta,\phi)
\odot
\mathbf e_m(\theta,\phi)
\nonumber\\
&=
\operatorname{diag}\{\mathbf e_m(\theta,\phi)\}
\mathbf a_{\rm UPA}(\theta,\phi),
\label{eq:fim_steering}
\end{align}
where $\odot$ denotes the Hadamard product.

The first morphing configuration is selected as the undeformed
reference state, i.e.,
\begin{equation}
\mathbf y_1=\mathbf 0_N,
\qquad
\mathbf e_1(\theta,\phi)=\mathbf 1_N .
\label{eq:reference_configuration}
\end{equation}
This reference configuration will later be used to resolve the
scaling ambiguities between the spatial and morphing factors.
During uplink training, user $k$ transmits the known pilot sequence
\begin{equation}
\mathbf s_k
=
[s_{k,1},\ldots,s_{k,T}]^T
\in\mathbb C^T ,
\label{eq:user_pilot}
\end{equation}
where $T$ denotes the number of pilot symbols transmitted under each
morphing configuration. {\color{blue}The pilot sequences are collected in the pilot matrix}
\begin{equation}
\mathbf S
=
[\mathbf s_1,\ldots,\mathbf s_K]^T
\in\mathbb C^{K\times T}.
\label{eq:pilot_matrix}
\end{equation}

Let $\alpha_k\in\mathbb C$ denote the complex channel gain of user
$k$, including its propagation attenuation and phase, and let
$(\theta_k,\phi_k)$ denote its angular parameters. Under the
single-dominant-path model, the channel vector between user $k$ and
the FIM under the $m$th morphing configuration is
\begin{align}
\mathbf h_{k,m}
&=
\alpha_k\mathbf a_m(\theta_k,\phi_k)
\nonumber\\
&=
\alpha_k
\operatorname{diag}\{\mathbf e_m(\theta_k,\phi_k)\}
\mathbf a_{\rm UPA}(\theta_k,\phi_k).
\label{eq:user_channel}
\end{align}
Collecting the element-domain observations over the $T$ pilot symbols
under the $m$th morphing configuration yields
\begin{align}
\mathbf Y_m
&=
\sum_{k=1}^{K}
\mathbf h_{k,m}\mathbf s_k^T+\mathbf W_m
\nonumber\\
&=
\mathbf H_m\mathbf S+\mathbf W_m,
\qquad m=1,\ldots,M,
\label{eq:received_matrix}
\end{align}
where
\begin{equation}
\mathbf H_m
=
[\mathbf h_{1,m},\ldots,\mathbf h_{K,m}]
\in\mathbb C^{N\times K}
\end{equation}
is the channel matrix under the $m$th morphing configuration, and
$\mathbf W_m\in\mathbb C^{N\times T}$ denotes additive noise whose
entries are independently distributed as
$\mathcal{CN}(0,\sigma_w^2)$.

The physical channel parameters are assumed to remain unchanged over
all pilot observations collected across the $M$ morphing configurations.
Let $T_s$ denote the pilot-symbol duration, $T_{\rm sw}$ the switching
time between two consecutive morphing states, and $T_c$ the channel
coherence time. The adopted block-fading model therefore requires
\begin{equation}
MTT_s+(M-1)T_{\rm sw}\leq T_c .
\label{eq:coherence_requirement}
\end{equation}
Under this condition, the observations acquired under different
morphing configurations can be jointly processed.

\subsection{Tensor Observation for Angularly Separated Users}
\label{subsec:tensor_observation}

The received matrices $\{\mathbf Y_m\}_{m=1}^{M}$ are stacked along
the morphing-configuration dimension to form the third-order
observation tensor
\begin{equation}
\mathcal Y
\in\mathbb C^{N\times T\times M},
\qquad
\mathcal Y(:,:,m)=\mathbf Y_m,
\quad m=1,\ldots,M.
\label{eq:observation_tensor}
\end{equation}
Its three modes correspond to the FIM elements, pilot symbols, and
morphing configurations, respectively. The corresponding noise tensor
$\mathcal W\in\mathbb C^{N\times T\times M}$ is constructed in the
same manner from $\{\mathbf W_m\}_{m=1}^{M}$.

For the structured FIM-BTD formulation developed in the next section,
it is convenient to interchange the pilot and morphing modes. We
therefore define
\begin{equation}
\widetilde{\mathcal Y}
=
\operatorname{permute}(\mathcal Y,[1,3,2])
\in\mathbb C^{N\times M\times T},
\label{eq:permuted_tensor}
\end{equation}
or equivalently,
\begin{equation}
[\widetilde{\mathcal Y}]_{n,m,t}
=
[\mathcal Y]_{n,t,m}.
\end{equation}
The corresponding permuted noise tensor is denoted by
$\widetilde{\mathcal W}$. Clearly, $\mathcal Y$ and
$\widetilde{\mathcal Y}$ contain the same measurements and differ
only in their mode ordering.

We focus on the angularly separated-user configuration, where the
users generate distinct angular responses,
\begin{equation}
(\theta_k,\phi_k)
\neq
(\theta_\ell,\phi_\ell),
\qquad
k\neq \ell.
\label{eq:angularly_separated}
\end{equation}
Hence, each user is associated with its own reference-array response
and morphing response. In particular, defining
\begin{align}
\mathbf a_k&
\triangleq
\mathbf a_{\rm UPA}(\theta_k,\phi_k),\\
\mathbf e_{k,m}&
\triangleq
\mathbf e_m(\theta_k,\phi_k),
\end{align}
the received observations contain $K$ user-specific
spatial--morphing contributions. %rather than a single common response.
This structure motivates the block-term representation developed in
Section~\ref{sec:btd_model}.

The physical parameters to be recovered are collected as
\begin{equation}
\boldsymbol{\theta}
=
[\theta_1,\ldots,\theta_K]^T,\qquad
\boldsymbol{\phi}
=
[\phi_1,\ldots,\phi_K]^T,
\end{equation}
and
\begin{equation}
\boldsymbol{\alpha}
=
[\alpha_1,\ldots,\alpha_K]^T.
\end{equation}
Given the observation tensor, the known pilot matrix $\mathbf S$,
and the prescribed morphing profiles
$\{\mathbf y_m\}_{m=1}^{M}$, the objective is to jointly estimate
$\{\boldsymbol{\theta},\boldsymbol{\phi},
\boldsymbol{\alpha}\}$ and reconstruct the configuration-dependent
channels $\{\mathbf h_{k,m}\}$.

\textit{Remark 1:}
If $\mathbf S$ has full row rank, the channel under each morphing
configuration can be independently estimated through
\begin{equation}
\widehat{\mathbf H}_m
=
\mathbf Y_m\mathbf S^\dagger.
\label{eq:separate_ls}
\end{equation}
{\color{blue}Here, $(\cdot)^\dagger$ denotes the Moore--Penrose pseudoinverse.}
However, such a configuration-wise least-squares approach processes
the morphing states independently and does not exploit their common
geometric structure. The structured FIM-BTD formulation developed
next instead jointly processes all morphing configurations while
preserving the known pilot directions and the user-specific
spatial--morphing structure.

\section{Structured FIM-BTD Formulation}
\label{sec:btd_model}

We now derive the structured BTD for the angularly separated-user setting in \eqref{eq:angularly_separated}. Since each user has a distinct reference-array response and morphing-dependent phase profile, the FIM-element and morphing modes are coupled through a user-specific matrix. The resulting observation is therefore a sum of $K$ blocks. %rather than a decomposition with one shared spatial factor.

\subsection{User-Specific Block-Term Representation}
\label{subsec:btd_representation}

We use the permuted observation tensor
$\widetilde{\mathcal Y}\in\mathbb C^{N\times M\times T}$
defined in \eqref{eq:permuted_tensor}, whose three modes correspond
to the FIM elements, morphing configurations, and pilot symbols,
respectively.
For user $k$, define the undeformed reference-array response as
\begin{equation}
\mathbf a_k
\triangleq
\mathbf a_{\rm UPA}(\theta_k,\phi_k)
\in\mathbb C^{N},
\label{eq:btd_ak}
\end{equation}
and construct the corresponding diagonal spatial factor
\begin{equation}
\mathbf A_k
\triangleq
\operatorname{diag}\{\mathbf a_k\}
\in\mathbb C^{N\times N}.
\label{eq:btd_Ak}
\end{equation}
The morphing responses of user $k$ over the $M$ prescribed
configurations are collected as
\begin{equation}
\mathbf E_k
\triangleq
\left[
\mathbf e_1(\theta_k,\phi_k),
\ldots,
\mathbf e_M(\theta_k,\phi_k)
\right]^T
\in\mathbb C^{M\times N}.
\label{eq:btd_Ek}
\end{equation}
{\color{blue}During factor recovery, $\mathbf E_k$ is treated as an unknown matrix, while its dependence on $(\theta_k,\phi_k)$ is enforced by
matching the recovered factor to the calibrated FIM manifold.}
Using \eqref{eq:user_channel}, the contribution of user $k$ to the
$t$th frontal slice of $\widetilde{\mathcal Y}$ is
\begin{equation}
\widetilde{\mathbf X}_{k,t}
=
\alpha_k s_{k,t}
\mathbf A_k\mathbf E_k^T
\in\mathbb C^{N\times M}.
\label{eq:btd_user_slice}
\end{equation}
Stacking these slices over the $T$ pilot symbols yields the
user-specific tensor block
\begin{equation}
\widetilde{\mathcal X}_k
=
\left(
\mathbf A_k\mathbf E_k^T
\right)
\circ
\left(
\alpha_k\mathbf s_k
\right),
\label{eq:btd_user_block}
\end{equation}
{\color{blue}where $\circ$ denotes the outer product between a matrix and a
vector. Specifically, $(\mathbf B\circ\mathbf c)_{n,m,t}
=b_{n,m}c_t$ for $\mathbf B\in\mathbb C^{N\times M}$ and
$\mathbf c\in\mathbb C^T$.}
The complete observation tensor can therefore be expressed as
\begin{equation}
\widetilde{\mathcal Y}
=
\sum_{k=1}^{K}
\left(
\mathbf A_k\mathbf E_k^T
\right)
\circ
\left(
\alpha_k\mathbf s_k
\right)
+
\widetilde{\mathcal W}.
\label{eq:structured_btd_model}
\end{equation}

Equation~\eqref{eq:structured_btd_model} defines the structured FIM-BTD model. Unlike a generic BTD, the known pilot fixes the direction of each block in the third mode, while the first-mode factor is constrained to be diagonal and therefore has only $N$ free spatial coefficients. The resulting model embeds user-to-pilot association and the elementwise FIM response directly in the factorization.

\subsection{Block Multilinear Rank}
\label{subsec:btd_rank}

The multilinear rank of each user block is determined by its
morphing factor. Define
\begin{equation}
r_{E,k}
\triangleq
\operatorname{rank}(\mathbf E_k)
\leq
\min\{M,N\}.
\label{eq:morphing_rank}
\end{equation}
Under the ideal steering model, all entries of $\mathbf a_k$ are
nonzero, such that $\mathbf A_k$ is nonsingular. Therefore,
multiplication by $\mathbf A_k$ does not alter the rank of
$\mathbf E_k^T$, and the $k$th tensor block satisfies
\begin{equation}
\operatorname{rank}_{\rm ML}
\left(
\widetilde{\mathcal X}_k
\right)
=
(r_{E,k},\,r_{E,k},\,1).
\label{eq:btd_multilinear_rank}
\end{equation}
Here, $\operatorname{rank}_{\rm ML}$ denotes the tuple of matrix
ranks of the three mode unfoldings of the tensor block.
Accordingly, the frequently used rank-$(N,N,1)$ description is valid
only when
\begin{equation}
M\geq N,
\qquad
\operatorname{rank}(\mathbf E_k)=N.
\label{eq:full_btd_rank_condition}
\end{equation}
When $M<N$ or the morphing responses are rank deficient, the same
structured block representation remains valid, but the corresponding
multilinear rank is reduced to
$(r_{E,k},r_{E,k},1)$.

\subsection{Structured Matrix Representations}
\label{subsec:btd_unfoldings}

{\color{blue}We use column-major vectorization. For clarity, the mode-2 and mode-3
unfoldings are defined by
\begin{align*}
\bigl([\widetilde{\mathcal Y}]_{(2)}\bigr)_{m,\,n+N(t-1)}
&=[\widetilde{\mathcal Y}]_{n,m,t},\\
\bigl([\widetilde{\mathcal Y}]_{(3)}\bigr)_{t,\,n+N(m-1)}
&=[\widetilde{\mathcal Y}]_{n,m,t},
\end{align*}
for $n=1,\ldots,N$, $m=1,\ldots,M$, and $t=1,\ldots,T$. The noise unfoldings follow the same convention. We next express \eqref{eq:structured_btd_model} in several complementary matrix forms.}

First, collect the user-specific morphing factors as
\begin{equation}
\mathbf E_{\rm B}
\triangleq
\left[
\mathbf E_1,\ldots,\mathbf E_K
\right]
\in\mathbb C^{M\times KN},
\label{eq:EB}
\end{equation}
and define the gain-weighted pilot matrix
\begin{equation}
\mathbf S(\boldsymbol{\alpha})
\triangleq
\mathbf S^T
\operatorname{diag}\{\boldsymbol{\alpha}\}
\in\mathbb C^{T\times K},
\label{eq:gain_weighted_pilot}
\end{equation}
whose $k$th column is $\alpha_k\mathbf s_k$.
Under the adopted unfolding convention, the mode-2 matricization of
\eqref{eq:structured_btd_model} can be written as
\begin{equation}
[\widetilde{\mathcal Y}]_{(2)}
=
\mathbf E_{\rm B}\mathbf Z_E^T
+
[\widetilde{\mathcal W}]_{(2)},
\label{eq:btd_mode2_E}
\end{equation}
where
\begin{equation}
\mathbf Z_E
\triangleq
\left[
\alpha_1\mathbf s_1\otimes\mathbf A_1,
\ldots,
\alpha_K\mathbf s_K\otimes\mathbf A_K
\right]
\in\mathbb C^{TN\times KN}.
\label{eq:ZE}
\end{equation}
Equation~\eqref{eq:btd_mode2_E} provides a linear model for estimating
the concatenated morphing factor $\mathbf E_{\rm B}$ when the spatial
factors and user gains are fixed.

To explicitly exploit the diagonal structure of
$\mathbf A_k$, stack the user-specific steering vectors as
\begin{equation}
\mathbf a_{\rm EXT}
\triangleq
\left[
\mathbf a_1^T,\ldots,\mathbf a_K^T
\right]^T
\in\mathbb C^{KN}.
\label{eq:a_ext}
\end{equation}
Vectorizing \eqref{eq:btd_mode2_E} gives
\begin{equation}
\operatorname{vec}
\left(
[\widetilde{\mathcal Y}]_{(2)}
\right)
=
\boldsymbol{\Phi}_a
\mathbf a_{\rm EXT}
+
\operatorname{vec}
\left(
[\widetilde{\mathcal W}]_{(2)}
\right),
\label{eq:btd_spatial_linear_model}
\end{equation}
where
\begin{equation}
\boldsymbol{\Phi}_a
\triangleq
\left(
\mathbf S(\boldsymbol{\alpha})
\otimes
\mathbf I_N
\right)
\diamond
\mathbf E_{\rm B}
\in\mathbb C^{MNT\times KN},
\label{eq:Phi_a}
\end{equation}

{\color{blue}Here, $\diamond$ denotes the columnwise Kronecker product, also known
as the Khatri--Rao product. From the representation in
\eqref{eq:btd_spatial_linear_model}, it is worth noting that instead of treating each
$\mathbf A_k\in\mathbb C^{N\times N}$ as an unconstrained matrix,
only its $N$ diagonal coefficients need to be estimated. Therefore,
the number of spatial unknowns is reduced from $KN^2$ for an unconstrained block model to only $KN$ coefficients in the proposed structured formulation.}
% \begin{equation}
% KN^2
% \end{equation}
% for an unconstrained block model to only
% \begin{equation}
% KN
% \end{equation}
% coefficients in the proposed structured formulation.

Finally, define the effective spatial--morphing response of user $k$
as
\begin{equation}
\mathbf g_k
\triangleq
\operatorname{vec}
\left(
\mathbf A_k\mathbf E_k^T
\right)
\in\mathbb C^{MN},
\label{eq:gk}
\end{equation}
and collect them as
\begin{equation}
\mathbf G
\triangleq
[\mathbf g_1,\ldots,\mathbf g_K]
\in\mathbb C^{MN\times K}.
\label{eq:G_matrix}
\end{equation}
The mode-3 unfolding of
\eqref{eq:structured_btd_model} then becomes
\begin{equation}
[\widetilde{\mathcal Y}]_{(3)}
=
\mathbf S^T
\operatorname{diag}\{\boldsymbol{\alpha}\}
\mathbf G^T
+
[\widetilde{\mathcal W}]_{(3)}.
\label{eq:btd_mode3}
\end{equation}

Equations~\eqref{eq:btd_mode2_E},
\eqref{eq:btd_spatial_linear_model}, and
\eqref{eq:btd_mode3} provide three complementary linear
representations for the unknown morphing factors, diagonal spatial
factors, and user gains, respectively. These structures form the basis
of the pilot-constrained structured FIM-BTD estimator developed in
the next section.

%\section{Pilot-Constrained Structured\\ FIM-BTD Estimator}
\section{Proposed Structured FIM-BTD Estimator}
\label{sec:btd_estimator}

{\color{blue}Building on the complementary matrix representations in Section~\ref{sec:btd_model}, we estimate the user-specific morphing factors, diagonal spatial factors, and complex gains by alternating structured least-squares updates. Unlike a generic BTD algorithm, the proposed updates preserve the known pilot directions and enforce $\mathbf A_k=\operatorname{diag}\{\mathbf a_k\}$ at every factor-recovery step. 
%The equations below give the unregularized least-squares updates; the numerical implementation uses Tikhonov regularization, as specified in the simulation setup.
}

\subsection{Alternating Structured Factor Updates}
\label{subsec:btd_factor_updates}

Let $\{\mathbf A_k^{(r-1)}\}_{k=1}^{K}$ and
$\boldsymbol{\alpha}^{(r-1)}$ denote the spatial-factor and gain
estimates available at the beginning of the $r$th iteration.

Using the current estimates in \eqref{eq:ZE}, define
\begin{equation}
\mathbf Z_E^{(r-1)}
=
\left[
\alpha_1^{(r-1)}
\mathbf s_1\otimes\mathbf A_1^{(r-1)},
\ldots,
\alpha_K^{(r-1)}
\mathbf s_K\otimes\mathbf A_K^{(r-1)}
\right].
\label{eq:ZE_iteration}
\end{equation}
Holding $\mathbf Z_E^{(r-1)}$ fixed, the morphing-factor update
follows from \eqref{eq:btd_mode2_E} as
\begin{equation}
\widetilde{\mathbf E}_{\rm B}^{(r)}
=
[\widetilde{\mathcal Y}]_{(2)}
\left[
\left(
\mathbf Z_E^{(r-1)}
\right)^T
\right]^{\dagger},
\label{eq:btd_E_update}
\end{equation}
where $(\cdot)^{\dagger}$ denotes the Moore--Penrose pseudoinverse.

Next, the updated morphing factor is inserted into the spatial
regression model. Retaining the previous gain estimate, define
\begin{equation}
\boldsymbol{\Phi}_a^{(r)}
=
\left[
\left(
\mathbf S^T
\operatorname{diag}
\{\boldsymbol{\alpha}^{(r-1)}\}
\otimes
\mathbf I_N
\right)
\diamond
\widetilde{\mathbf E}_{\rm B}^{(r)}
\right].
\label{eq:Phi_a_iteration}
\end{equation}
The extended steering vector is then updated through
\begin{equation}
\widetilde{\mathbf a}_{\rm EXT}^{(r)}
=
\left(
\boldsymbol{\Phi}_a^{(r)}
\right)^{\dagger}
\operatorname{vec}
\left(
[\widetilde{\mathcal Y}]_{(2)}
\right).
\label{eq:btd_a_update}
\end{equation}

The update in \eqref{eq:btd_a_update} directly estimates only the
$KN$ diagonal coefficients of the $K$ spatial factors, instead of
the $KN^2$ entries required by an unconstrained block model.

\subsection{Reference-State Normalization}
\label{subsec:btd_normalization}

The factorization in
\eqref{eq:structured_btd_model} contains reciprocal scaling
ambiguities between the spatial and morphing factors. These ambiguities
are removed by exploiting the undeformed reference configuration.

Partition
$\widetilde{\mathbf E}_{\rm B}^{(r)}$ and
$\widetilde{\mathbf a}_{\rm EXT}^{(r)}$ into their user-specific
components
\begin{equation}
\widetilde{\mathbf E}_{\rm B}^{(r)}
=
\left[
\widetilde{\mathbf E}_1^{(r)},
\ldots,
\widetilde{\mathbf E}_K^{(r)}
\right]
\end{equation}
and
\begin{equation}
\widetilde{\mathbf a}_{\rm EXT}^{(r)}
=
\left[
(\widetilde{\mathbf a}_1^{(r)})^T,
\ldots,
(\widetilde{\mathbf a}_K^{(r)})^T
\right]^T .
\end{equation}
For user $k$, define
\begin{equation}
\mathbf D_{E,k}^{(r)}
\triangleq
\operatorname{diag}
\left\{
\left(
\widetilde{\mathbf E}_k^{(r)}(1,:)
\right)^T
\right\}.
\label{eq:DEk}
\end{equation}
{\color{blue}Normalization requires nonzero entries in the estimated reference
row.}
Using the reference-state condition
$\mathbf E_k(1,:)=\mathbf 1_N^T$, normalize the two factors as
\begin{align}
\mathbf E_k^{(r)}
&=
\widetilde{\mathbf E}_k^{(r)}
\left(
\mathbf D_{E,k}^{(r)}
\right)^{-1},
\nonumber\\
\mathbf a_k^{(r)}
&=
\mathbf D_{E,k}^{(r)}
\widetilde{\mathbf a}_k^{(r)}.
\label{eq:btd_reference_normalization}
\end{align}
This reciprocal rescaling preserves
$\mathbf A_k\mathbf E_k^T$ while enforcing
\begin{equation}
\mathbf E_k^{(r)}(1,:)
=
\mathbf 1_N^T.
\end{equation}

A residual scalar ambiguity remains between
$\mathbf a_k^{(r)}$ and $\alpha_k$. Since the first entry of the
reference UPA steering vector is normalized to unity, define
\begin{equation}
\gamma_k^{(r)}
\triangleq
[\mathbf a_k^{(r)}]_1 .
\label{eq:gamma_k}
\end{equation}
The spatial factor is normalized according to
\begin{equation}
\mathbf a_k^{(r)}
\leftarrow
\frac{\mathbf a_k^{(r)}}{\gamma_k^{(r)}},
\qquad
\mathbf A_k^{(r)}
=
\operatorname{diag}
\left\{
\mathbf a_k^{(r)}
\right\}.
\label{eq:btd_spatial_normalization}
\end{equation}
The corresponding block amplitude and phase are subsequently absorbed
into the channel gain update. Under the ideal steering model,
$\gamma_k^{(r)}$ is nonzero in the noise-free case. In our numerical
experiments, a small threshold $\delta_{\gamma}>0$ is used.
If
\begin{equation}
|\gamma_k^{(r)}|<\delta_{\gamma},
\end{equation}
the current normalized factor pair for user $k$ is discarded and the
previous valid pair is retained to avoid unstable scaling.

\subsection{Direct User-Gain Update}
\label{subsec:btd_gain_update}

After normalization, define the effective spatial--morphing response
of user $k$ as
\begin{equation}
\mathbf g_k^{(r)}
=
\operatorname{vec}
\left[
\mathbf A_k^{(r)}
\left(
\mathbf E_k^{(r)}
\right)^T
\right],
\label{eq:gk_iteration}
\end{equation}
and collect these responses into
\begin{equation}
\mathbf G^{(r)}
=
\left[
\mathbf g_1^{(r)},
\ldots,
\mathbf g_K^{(r)}
\right].
\label{eq:G_iteration}
\end{equation}
Using the mode-3 representation in \eqref{eq:btd_mode3},
vectorization yields the linear gain model
\begin{equation}
\operatorname{vec}
\left(
[\widetilde{\mathcal Y}]_{(3)}
\right)
=
\boldsymbol{\Phi}_{\alpha}^{(r)}
\boldsymbol{\alpha}
+
\operatorname{vec}
\left(
[\widetilde{\mathcal W}]_{(3)}
\right),
\end{equation}
where
\begin{equation}
\boldsymbol{\Phi}_{\alpha}^{(r)}
\triangleq
\mathbf G^{(r)}
\diamond
\mathbf S^T
\in
\mathbb C^{MNT\times K}.
\label{eq:Phi_alpha}
\end{equation}
The user gains are therefore directly updated as
\begin{equation}
\boldsymbol{\alpha}^{(r)}
=
\left(
\boldsymbol{\Phi}_{\alpha}^{(r)}
\right)^{\dagger}
\operatorname{vec}
\left(
[\widetilde{\mathcal Y}]_{(3)}
\right).
\label{eq:btd_alpha_update}
\end{equation}

An important feature of
\eqref{eq:btd_alpha_update} is that the known pilot vectors remain
explicitly embedded in $\boldsymbol{\Phi}_{\alpha}^{(r)}$. Hence,
the recovered blocks remain associated with their corresponding users,
and an additional block-permutation matching stage is not required.

\subsection{Residual Criterion and Initialization}
\label{subsec:btd_convergence}

After each complete factor update, the normalized reconstruction
residual is evaluated as
\begin{equation}
\xi_{\rm B}^{(r)}
=
\frac{
	\left\|
	\widetilde{\mathcal Y}
	-
	\displaystyle\sum_{k=1}^{K}
	\left[
	\mathbf A_k^{(r)}
	\left(
	\mathbf E_k^{(r)}
	\right)^T
	\right]
	\circ
	\left(
	\alpha_k^{(r)}\mathbf s_k
	\right)
	\right\|_F^2
}{
	\left\|
	\widetilde{\mathcal Y}
	\right\|_F^2
}.
\label{eq:btd_residual}
\end{equation}

The alternating updates terminate when
\begin{equation}
\frac{
	\left|
	\xi_{\rm B}^{(r)}
	-
	\xi_{\rm B}^{(r-1)}
	\right|
}{
	\max
	\left\{
	\xi_{\rm B}^{(r-1)},
	\epsilon_0
	\right\}
}
\leq
\varepsilon_{\rm B},
\label{eq:btd_stopping}
\end{equation}
where $\varepsilon_{\rm B}>0$ denotes the convergence tolerance and
$\epsilon_0>0$ prevents division by zero. The iterations are also
terminated when the prescribed maximum number of iterations,
$I_{\rm B}$, is reached. We initialize the estimator from pilot-domain
least-squares channel estimates. Specifically, for each configuration,
\begin{equation}
\widehat{\mathbf H}_m^{\rm LS}
=
\mathbf Y_m\mathbf S^{\dagger}.
\label{eq:btd_initial_ls}
\end{equation}
The undeformed reference configuration initializes the user-specific
reference-array responses and gains, while the remaining
pilot-separated channel estimates initialize the morphing factors.

Since the alternating optimization is nonconvex, monotonic convergence
to the global optimum is not guaranteed. We therefore retain the
factor set producing the smallest reconstruction residual encountered
during the iterations.

\begin{algorithm}[t]
	\caption{Proposed Structured FIM-BTD Estimator}
	\label{alg:btd}
	\begin{algorithmic}[1]
	\color{blue}
		
\Require Observation tensor $\widetilde{\mathcal Y}$, known pilot matrix $\mathbf S$, initial factors, $\varepsilon_{\rm B}$, $\delta_{\gamma}$, and $I_{\rm B}$
		\Ensure $\{\widehat{\mathbf a}_k,\widehat{\mathbf E}_k,\widehat{\alpha}_k\}_{k=1}^{K}$
		
		\State Evaluate the initial residual $\xi_{\rm B}^{(0)}$
		\State Set $\xi_{\rm B}^{\star}\leftarrow\xi_{\rm B}^{(0)}$ and store the initial factor set
		\State Set $r\leftarrow1$
		
\While{$r\leq I_{\rm B}$}
		
		\State Construct $\mathbf Z_E^{(r-1)}$ using \eqref{eq:ZE_iteration}
		\State Update $\widetilde{\mathbf E}_{\rm B}^{(r)}$ using \eqref{eq:btd_E_update}
		
		\State Construct $\boldsymbol{\Phi}_a^{(r)}$ using \eqref{eq:Phi_a_iteration}
		\State Update $\widetilde{\mathbf a}_{\rm EXT}^{(r)}$ using \eqref{eq:btd_a_update}
		
		\State Normalize $\mathbf E_k^{(r)}$ and $\mathbf a_k^{(r)}$ using
		\eqref{eq:btd_reference_normalization} and \eqref{eq:btd_spatial_normalization}
		
		\State If $|\gamma_k^{(r)}|<\delta_{\gamma}$ for any user, retain its previous valid normalized factor pair
		
		\State Construct $\mathbf G^{(r)}$ using \eqref{eq:G_iteration}
		\State Update $\boldsymbol{\alpha}^{(r)}$ using \eqref{eq:btd_alpha_update}
		
		\State Evaluate $\xi_{\rm B}^{(r)}$ using \eqref{eq:btd_residual}
		
		\If{$\xi_{\rm B}^{(r)}<\xi_{\rm B}^{\star}$}
		\State Set $\xi_{\rm B}^{\star}\leftarrow\xi_{\rm B}^{(r)}$ and store the current factor set
		\EndIf
		
		\If{the stopping criterion in \eqref{eq:btd_stopping} is satisfied}
		\State \textbf{break}
		\EndIf
		
		\State $r\leftarrow r+1$
		
		\EndWhile
		
		\State \textbf{return} the stored minimum-residual factor set
		
	\color{black}
	\end{algorithmic}
\end{algorithm}

\subsection{Physical Parameter and Channel Recovery}
\label{subsec:btd_parameter_recovery}

Let
$\{\widehat{\mathbf a}_k,
\widehat{\mathbf E}_k,
\widehat{\alpha}_k\}_{k=1}^{K}$
denote the factor estimates returned by
Algorithm~\ref{alg:btd}. Since the known pilot directions are
preserved throughout the factorization, each estimated block is already
associated with its corresponding user.

Define the calibrated morphing matrix associated with an arbitrary
angular pair as
\begin{equation}
\mathbf E(\theta,\phi)
=
\left[
\mathbf e_1(\theta,\phi),
\ldots,
\mathbf e_M(\theta,\phi)
\right]^T.
\label{eq:btd_manifold_E}
\end{equation}
The angular parameters of user $k$ are extracted by jointly matching
its recovered spatial and morphing factors to the calibrated FIM
manifold:
\begin{align}
(\widehat{\theta}_k,\widehat{\phi}_k)
=
\arg\min_{(\theta,\phi)\in\Omega}
\;
&
1-
\frac{
	\left|
	\mathbf a_{\rm UPA}^{H}(\theta,\phi)
	\widehat{\mathbf a}_k
	\right|^2
}{
	\left\|
	\mathbf a_{\rm UPA}(\theta,\phi)
	\right\|_2^2
	\left\|
	\widehat{\mathbf a}_k
	\right\|_2^2
}
\nonumber\\
&+
\lambda_E
\frac{
	\left\|
	\widehat{\mathbf E}_k
	-
	\mathbf E(\theta,\phi)
	\right\|_F^2
}{
	\left\|
	\widehat{\mathbf E}_k
	\right\|_F^2
},
\label{eq:btd_angle_recovery}
\end{align}
{\color{blue}where $\Omega$ denotes the admissible angular region and
$\lambda_E\geq0$ weights the morphing-factor mismatch relative to the
spatial steering-vector mismatch.}

Finally, the configuration-dependent channel of user $k$ under the
$m$th morphing configuration is reconstructed as
\begin{equation}
\widehat{\mathbf h}_{k,m}
=
\widehat{\alpha}_k
\mathbf a_m
\left(
\widehat{\theta}_k,
\widehat{\phi}_k
\right),
\qquad
k=1,\ldots,K,\;
m=1,\ldots,M.
\label{eq:btd_channel_reconstruction}
\end{equation}

{\color{blue}The diagonal constraint
$\mathbf A_k=\operatorname{diag}\{\mathbf a_k\}$ reduces the number
of user-specific spatial unknowns from $KN^2$ to $KN$. The
undeformed reference configuration fixes the columnwise scaling between
$\mathbf A_k$ and $\mathbf E_k$, while the normalization
$[\mathbf a_k]_1=1$ assigns the remaining block amplitude and phase to
$\alpha_k$. These conventions fix factor scaling but do not, by
themselves, guarantee global uniqueness. Stable recovery also requires
sufficiently distinct user-specific spatial--morphing responses and
well-conditioned pilot sequences. The corresponding conditions are
discussed in the next section.}

\section{Identifiability and Computational Analysis}
\label{sec:identifiability}

{\color{blue}
This section specifies the rank conditions that make each conditional least-squares update unique, distinguishes these local conditions from global factor and physical-angle identifiability, and summarizes the pilot, morphing, coherence-time, and computational implications.

\subsection{Algebraic Well-Posedness and Physical Identifiability}
\label{subsec:btd_identifiability}

The proposed estimator alternates among three conditional least-squares updates for the morphing factors, diagonal spatial factors, and user gains. Their design matrices are $\mathbf Z_E$, $\boldsymbol{\Phi}_a$, and $\boldsymbol{\Phi}_{\alpha}$, respectively. The following full-column-rank conditions characterize uniqueness of the unregularized updates when the remaining factors are fixed. Note that a positive Tikhonov penalty can be used to make a regularized update unique even if its unregularized design matrix is rank deficient.
\begin{align}
\operatorname{rank}(\mathbf Z_E)&=KN,
\label{eq:rank_ZE}\\
\operatorname{rank}(\boldsymbol{\Phi}_a)&=KN,
\label{eq:rank_Phia}
\end{align}
and
\begin{equation}
\operatorname{rank}(\boldsymbol{\Phi}_{\alpha})=K.
\label{eq:rank_Phialpha}
\end{equation}
Since $\mathbf Z_E\in\mathbb C^{TN\times KN}$,
% \begin{equation}
% \mathbf Z_E\in\mathbb C^{TN\times KN},
% \end{equation}
a necessary dimensional condition to ensure 
\eqref{eq:rank_ZE} is
\begin{equation}
T\geq K.
\label{eq:T_geq_K}
\end{equation}
Similarly, since $\boldsymbol{\Phi}_a
\in \mathbb C^{MNT\times KN}$,
% \begin{equation}
% \boldsymbol{\Phi}_a
% \in
% \mathbb C^{MNT\times KN},
% \end{equation}
its full column-rank condition requires
\begin{equation}
MT\geq K.
\label{eq:MT_geq_K}
\end{equation}
These dimensional inequalities are necessary but not sufficient.
In particular, highly correlated pilots or nearly identical
user-specific spatial--morphing responses may render the design
matrices ill-conditioned and cause significant noise amplification.

It is important to distinguish the uniqueness of the individual
least-squares updates from the global uniqueness of the complete
structured BTD. Conditions
\eqref{eq:rank_ZE}--\eqref{eq:rank_Phialpha} guarantee that each
linear subproblem has a unique solution when the remaining factors
are fixed. They do not, by themselves, establish global uniqueness
of the complete nonconvex factorization or guarantee convergence of
the alternating procedure to the global optimum.
The structural constraints further reduce the ambiguities normally
associated with a generic block-term decomposition. In particular,
\begin{equation}
\mathbf A_k
=
\operatorname{diag}\{\mathbf a_k\}
\end{equation}
eliminates the general within-block transformation ambiguity since
the first-mode factor is restricted to a diagonal matrix. The
remaining columnwise scaling between $\mathbf A_k$ and $\mathbf E_k$
is removed through the undeformed-reference condition
\begin{equation}
\mathbf E_k(1,:)=\mathbf 1_N^T,
\label{eq:btd_ref_identifiability}
\end{equation}
whereas the residual block scaling is fixed by
\begin{equation}
[\mathbf a_k]_1=1.
\label{eq:btd_spatial_identifiability}
\end{equation}
These conventions assign the remaining block amplitude and phase to
$\alpha_k$. The known pilot vector $\mathbf s_k$ is explicitly
embedded in the $k$th block throughout the estimation procedure, so
the modeled blocks are associated with their users and no separate
block-permutation search is needed when the corresponding design
matrices have full column rank.

The available morphing diversity also determines the multilinear rank
of each user block. Recall from \eqref{eq:morphing_rank} that
\begin{equation}
r_{E,k}
=
\operatorname{rank}(\mathbf E_k).
\end{equation}
Since $\mathbf A_k$ is nonsingular under the ideal steering model,
\begin{equation}
\operatorname{rank}_{\rm ML}
\left(
\widetilde{\mathcal X}_k
\right)
=
(r_{E,k},r_{E,k},1).
\end{equation}
Therefore, a strict rank-$(N,N,1)$ block requires
\begin{equation}
M\geq N,
\qquad
\operatorname{rank}(\mathbf E_k)=N,
\quad
k=1,\ldots,K.
\label{eq:btd_strict_rank_condition}
\end{equation}
When this condition is not satisfied, the structured representation
remains valid with the reduced block rank
$(r_{E,k},r_{E,k},1)$.

Note that the above conditions are related to algebraic factor recovery. Unique recovery
of the physical angular parameters additionally depends on the FIM manifold. Define the joint spatial--morphing mapping
\begin{equation}
\mathcal G(\theta,\phi)
\triangleq
\left(
\mathbf a_{\rm UPA}(\theta,\phi),
\mathbf E(\theta,\phi)
\right),
\label{eq:btd_joint_manifold}
\end{equation}
where $\mathbf E(\theta,\phi)$ is defined in
\eqref{eq:btd_manifold_E}. Physical angle identification over the
admissible angular region $\Omega$ requires this mapping to be
injective, i.e.,
\begin{equation}
\mathcal G(\theta_1,\phi_1)
=
\mathcal G(\theta_2,\phi_2)
\Longrightarrow
(\theta_1,\phi_1)
=
(\theta_2,\phi_2),
\label{eq:btd_manifold_injective}
\end{equation}
for all
$(\theta_1,\phi_1),(\theta_2,\phi_2)\in\Omega$.

In addition to this single-user manifold identifiability, stable
multiuser recovery requires sufficiently distinct user-specific
spatial--morphing responses. 
%As two users approach the same angular
%direction, their corresponding blocks become increasingly coherent,
%leading to deteriorated conditioning and reduced numerical
%separability.

\subsection{Pilot and Morphing Requirements}
\label{subsec:btd_training_requirements}

The pilot and morphing dimensions provide complementary forms of
diversity. The pilot dimension separates user-dependent contributions,
whereas the morphing dimension provides geometric diversity for
recovering their configuration-dependent FIM responses.
A sufficient pilot-domain separation condition is
\begin{equation}
\operatorname{rank}(\mathbf S)=K,
\qquad
T\geq K.
\label{eq:btd_pilot_requirement}
\end{equation}
Equal-energy orthogonal pilots satisfy this condition when
\begin{equation}
\mathbf S\mathbf S^H
=
E_p\mathbf I_K,
\label{eq:orthogonal_pilots}
\end{equation}
where $E_p$ denotes the energy of each pilot sequence. 
The morphing configurations provide the second source of diversity.
The first configuration is selected as the undeformed reference,
while at least one additional configuration should contain nonzero
out-of-plane displacements:
\begin{equation}
\mathbf y_1=\mathbf 0_N,
\qquad
M\geq2,
\qquad
\sum_{m=2}^{M}
\|\mathbf y_m\|_2^2
>0.
\label{eq:btd_morphing_requirement}
\end{equation}
The undeformed state fixes the factor scaling, while nonzero morphing profiles introduce sensitivity to the directional component along the $y$-axis. %The joint manifold must also be injective over the angular region of interest. 
%Repeated or nearly identical displacement profiles provide little additional geometric diversity and can worsen the conditioning of the structured factorization.

For the proposed structured FIM-BTD estimator, a sufficient
pilot-separation condition and the principal conditional-update rank
requirements can be summarized as
\begin{align}
&\operatorname{rank}(\mathbf S)=K,\qquad T\geq K,
\nonumber\\
&\mathbf y_1=\mathbf 0_N,\qquad M\geq2,
\nonumber\\
&\sum_{m=2}^{M}\|\mathbf y_m\|_2^2>0,
\nonumber\\
&\operatorname{rank}(\mathbf Z_E)=KN,
\nonumber\\
&\operatorname{rank}(\boldsymbol{\Phi}_a)=KN,
\nonumber\\
&\operatorname{rank}(\boldsymbol{\Phi}_{\alpha})=K.
\label{eq:btd_summary_requirements}
\end{align}
A strict rank-$(N,N,1)$ characterization additionally requires
\eqref{eq:btd_strict_rank_condition}. Note that this stronger requirement is
not necessary for the validity of the structured BTD representation. For instance, when $\operatorname{rank}(\mathbf E_k)<N$, the corresponding block
simply possesses the reduced multilinear rank
$(r_{E,k},r_{E,k},1)$.
The total training overhead is
\begin{equation}
\tau_{\rm tr}=MT
\label{eq:btd_training_overhead}
\end{equation}
pilot-symbol intervals, resulting in $NMT$ complex scalar
observations. Increasing $T$ primarily improves pilot-domain user
separation and noise averaging, whereas increasing $M$ provides
additional morphing diversity. 
%Consequently, two choices of $(M,T)$
%having the same product $MT$ need not yield identical estimation
%performance.

Finally, all observations must be acquired within one channel
coherence interval. Combining the pilot-symbol duration $T_s$, the
morphing-state switching time $T_{\rm sw}$, and the coherence time
$T_c$, the block-fading assumption requires
\begin{equation}
MTT_s+(M-1)T_{\rm sw}
\leq
T_c.
\label{eq:btd_coherence_requirement}
\end{equation}

\subsection{Computational Complexity}
\label{subsec:btd_complexity}

We next characterize the dominant arithmetic complexity of the
proposed structured FIM-BTD estimator in terms of the number of FIM
elements $N$, morphing configurations $M$, pilot symbols $T$, users
$K$, and structured BTD iterations $I_{\rm B}$. Tensor permutation
and unfolding are regarded as index rearrangements and are therefore
not included in the dominant arithmetic count.

Each structured FIM-BTD iteration contains three principal
least-squares updates. The morphing-factor update in
\eqref{eq:btd_E_update} involves the coefficient matrix
$\mathbf Z_E^T\in\mathbb C^{KN\times TN}$. The diagonal spatial-factor
update uses
\begin{equation}
\boldsymbol{\Phi}_a
\in
\mathbb C^{MNT\times KN},
\end{equation}
while the direct gain update uses
\begin{equation}
\boldsymbol{\Phi}_{\alpha}
\in
\mathbb C^{MNT\times K}.
\end{equation}
With dense QR- or singular value decomposition (SVD)-based least-squares solvers, the structured
spatial update is the dominant operation. The resulting complexity of
the proposed estimator is
\begin{equation}
   C_{\rm BTD}
	=
	\mathcal O
	\left[
	I_{\rm B}
	\left(
	K^2MTN^3
	+
	K^3N^3
	+
	KMTN^2
	\right)
	\right].
\label{eq:btd_complexity}
\end{equation}
The direct gain update has the lower-order complexity
\begin{equation}
C_{\alpha}
=
\mathcal O
\left(
MNTK^2+K^3
\right).
\label{eq:btd_gain_complexity}
\end{equation}
Hence, for typical values of $N$, the overall computational burden is
mainly determined by the structured spatial-factor regression. % rather
%than the user-gain estimation.
For comparison, independently estimating the channel under each
morphing configuration through the pilot-domain least-squares
estimator
$\widehat{\mathbf H}_m=\mathbf Y_m\mathbf S^\dagger$
requires approximately
\begin{equation}
C_{\rm LS}
=
\mathcal O
\left(
MNTK+TK^2+K^3
\right).
\label{eq:pilot_ls_complexity}
\end{equation}
Although this configuration-wise LS estimator has substantially lower
computational complexity, it processes each morphing state
independently and does not exploit the geometric coupling across the
$M$ FIM configurations.
The matrix
$\boldsymbol{\Phi}_a\in\mathbb C^{MNT\times KN}$ can become large
for high-dimensional FIMs. Explicitly storing it requires
approximately
\begin{equation}
\mathcal O(MTK N^2)
\end{equation}
complex-valued entries. In practice, the associated products can be
implemented through matrix-free Kronecker and Khatri--Rao operations,
thereby avoiding explicit construction of this regression matrix.
If the physical angular parameters are extracted using a grid of
$N_{\Omega}$ candidate angle pairs, the additional manifold-search
complexity is
\begin{equation}
C_{\rm angle}
=
\mathcal O
\left(
KN_{\Omega}MN
\right).
\label{eq:btd_angle_complexity}
\end{equation}
This search cost is separate from the algebraic BTD factor-recovery
complexity in \eqref{eq:btd_complexity}.
}

\section{Simulation Results and Discussion}
\label{sec:simulation}

{\color{blue}
\subsection{Simulation Setup}
\label{subsec:simulation_setup}

We consider a sensing-enabled FIM composed of
$N=N_xN_z=16$ elements, where $N_x=N_z=4$. The undeformed
FIM lies on the $xz$-plane, and the inter-element spacings are set to
$d_x=d_z=\lambda/2$. Without loss of generality, the wavelength is
normalized as $\lambda=1$, yielding the wavenumber
$\kappa=2\pi/\lambda$. The FIM elements are displaced only along the
$y$-axis. The first configuration is selected as the undeformed
reference state, i.e., $\mathbf y_1=\mathbf 0_N$.
% \begin{equation}
% \mathbf y_1=\mathbf 0_N.
% \end{equation}
Unless otherwise specified, the displacement coefficients of the
remaining morphing configurations are independently generated from
$\mathcal U[-0.35\lambda,0.35\lambda]$.

The users transmit deterministic discrete Fourier transform (DFT)
pilot sequences. In the default setting, the pilot matrix satisfies
$\mathbf S\mathbf S^H=T\mathbf I_K$.
% \begin{equation}
% \mathbf S\mathbf S^H=T\mathbf I_K,
% \end{equation}. 
Unless otherwise stated, the
angularly separated-user scenario employs $K=3$, $T=12$, and  $M=16$.
% \begin{equation}
% K=3,\qquad T=12,\qquad M=16.
% \end{equation}
The elevation and azimuth angles are generated within
$\theta_k\in[25^\circ,70^\circ]$ and
$\phi_k\in[-55^\circ,55^\circ]$, respectively. To avoid nearly
coincident user directions in the default experiments, a realization
is accepted only when
\begin{equation}
\sqrt{
	(\theta_k-\theta_\ell)^2+
	(\phi_k-\phi_\ell)^2
}
\geq 15^\circ,
\qquad k\neq \ell.
\end{equation}
The user gains are generated as
\begin{equation}
\alpha_k=e^{j\psi_k},
\qquad
\psi_k\sim\mathcal U[0,2\pi).
\end{equation}
When the number of users is varied, the gains are normalized as
$\alpha_k=e^{j\psi_k}/\sqrt{K}$, such that $\sum_{k=1}^{K}|\alpha_k|^2=1$.
% \begin{equation}
% \sum_{k=1}^{K}|\alpha_k|^2=1.
% \end{equation}
For each Monte Carlo realization, let $\widetilde{\mathcal X}$ denote
the noiseless observation tensor obtained from
\eqref{eq:structured_btd_model} by setting the noise to zero. The
average signal power per observation entry is
\begin{equation}
P_X=
\frac{\|\widetilde{\mathcal X}\|_F^2}{NMT}.
\end{equation}
The noise entries are independently generated according to
$\mathcal{CN}(0,\sigma_w^2)$, where $\sigma_w^2
= P_X10^{-\mathrm{SNR}/10}$.
% \begin{equation}
% \sigma_w^2
% =
% P_X10^{-\mathrm{SNR}/10}.
% \end{equation}

The structured FIM-BTD estimator is initialized by the pilot-domain
least-squares estimates. The maximum number of BTD iterations is set
to $I_{\rm B}=20$, the stopping tolerance is $10^{-6}$, and the
Tikhonov regularization parameter is $10^{-6}$ for the least-squares
updates. After algebraic factor recovery, the physical
angles are initialized through a two-dimensional coarse search and
subsequently refined by continuous local optimization. The weighting
coefficient in the joint spatial--morphing manifold matching criterion
is set to $\lambda_E=1$.
Let $N_{\rm MC}$ denote the number of Monte Carlo realizations. The
final reconstructed-channel normalized mean squared error (NMSE) is
computed as
\begin{equation}
\mathrm{NMSE}_{\mathbf h}
=
\frac{1}{N_{\rm MC}}
\sum_{i=1}^{N_{\rm MC}}
\frac{
	\displaystyle
	\sum_{k=1}^{K}\sum_{m=1}^{M}
	\left\|
	\widehat{\mathbf h}_{k,m}^{(i)}
	-
	\mathbf h_{k,m}^{(i)}
	\right\|_2^2
}{
	\displaystyle
	\sum_{k=1}^{K}\sum_{m=1}^{M}
	\left\|
	\mathbf h_{k,m}^{(i)}
	\right\|_2^2
}.
\label{eq:sim_channel_nmse}
\end{equation}
%The corresponding value, in dB, is
%$10\log_{10}(\mathrm{NMSE}_{\mathbf h})$.

\begin{figure*}[t]
	\centering
	\subfloat[Final reconstructed-channel NMSE.]{
		\includegraphics[width=0.45\textwidth]
		{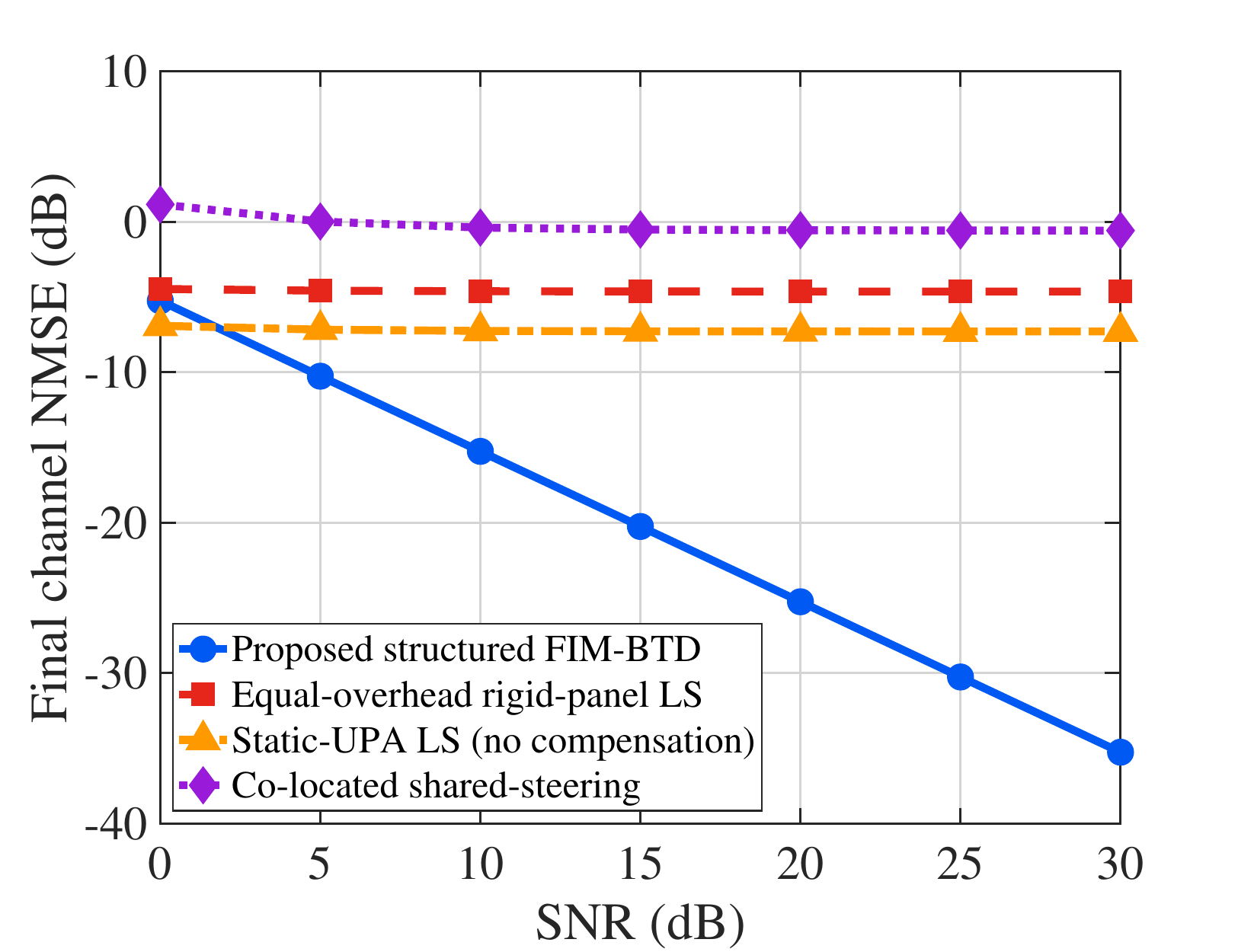}
		\label{fig:btd_snr_channel}
	}
	\hfill
	\subfloat[Joint angular RMSE.]{
		\includegraphics[width=0.48\textwidth]
		{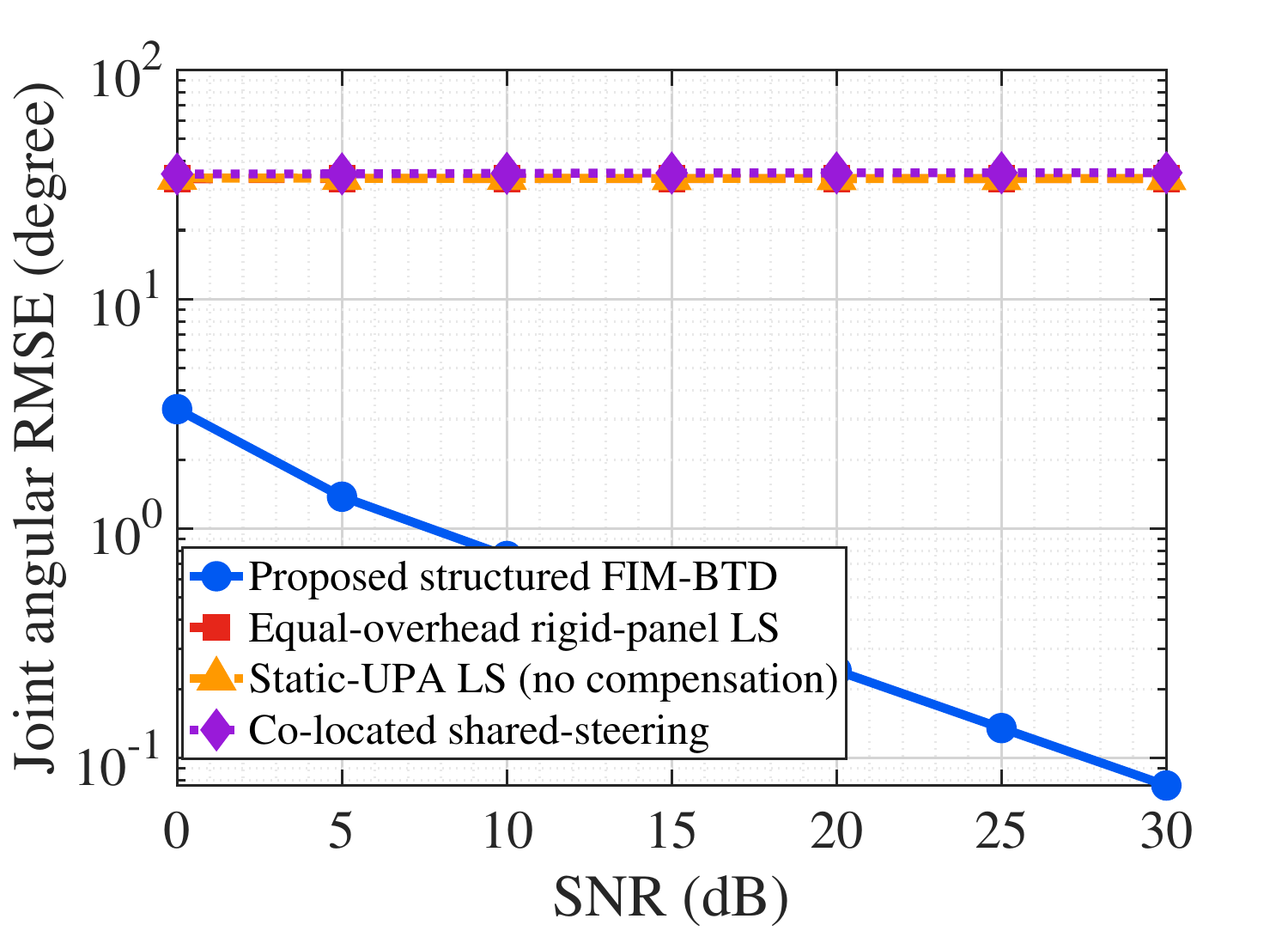}
		\label{fig:btd_snr_angle}
	}
	\caption{End-to-end estimation performance versus SNR for the
		angularly separated-user scenario with $N=16$, $K=3$, $T=12$,
		and $M=16$: (a) final reconstructed-channel NMSE;
		(b) joint angular RMSE.}
	\label{fig:btd_snr}
\end{figure*}

For angular estimation, let $\operatorname{wrap}_{[-180^\circ,180^\circ)}$
map an angular difference to its principal value in the indicated
interval. Define
$\Delta\theta_k^{(i)}
=\widehat{\theta}_k^{(i)}-\theta_k^{(i)}$
and
\begin{equation}
\Delta\phi_k^{(i)}
=
\operatorname{wrap}_{[-180^\circ,180^\circ)}
\left(
\widehat{\phi}_k^{(i)}-\phi_k^{(i)}
\right).
\end{equation}
The joint angular root mean square error (RMSE) is defined as
\begin{equation}
\mathrm{RMSE}_{\rm ang}
=
\sqrt{
	\frac{1}{2KN_{\rm MC}}
	\sum_{i=1}^{N_{\rm MC}}
	\sum_{k=1}^{K}
	\left[
	\left(\Delta\theta_k^{(i)}\right)^2
	+
	\left(\Delta\phi_k^{(i)}\right)^2
	\right]
}.
\label{eq:sim_angle_rmse}
\end{equation}
With both angles expressed in degrees, this RMSE is also in degrees.

For comparisons involving the rigid-panel least squares (LS) baseline, equal total
pilot-symbol overhead is imposed. Specifically, the proposed method
collects $T$ pilot symbols under each of the $M$ morphing
configurations, corresponding to $MT$ pilot-symbol intervals, whereas
the rigid-panel baseline remains in the undeformed configuration and
uses $T_{\rm rigid}=MT$ pilot symbols.

We compare the proposed method with three reference estimators: an
equal-overhead rigid-panel LS estimator; a static-UPA LS estimator
that ignores the morphing-dependent phase variation; and a
common-steering baseline that forces all users to share one spatial
response.

\begin{figure}[!t]
	\centering
	\includegraphics[width=.45\textwidth]
	{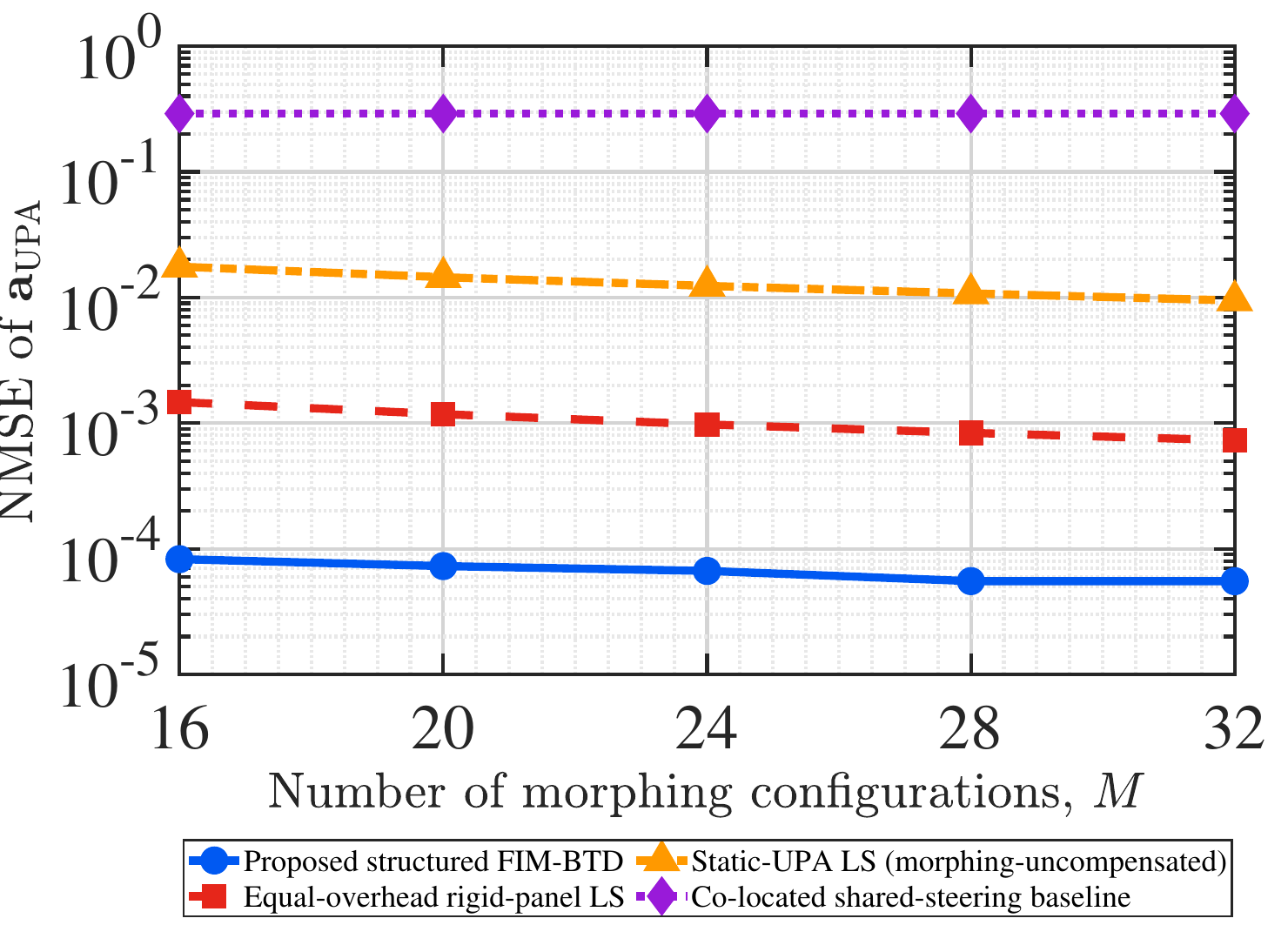}
	\caption{NMSE of the estimated user-specific UPA steering vectors
		versus the number of morphing configurations $M$ with $N=16$,
		$K=3$, $T=12$, and $\mathrm{SNR}=10$ dB.}
	\label{fig:btd_M}
\end{figure}

\subsection{Estimation Performance versus SNR}
\label{subsec:sim_snr}

Fig.~\ref{fig:btd_snr} compares end-to-end performance as the SNR
varies. The channel NMSE and angular RMSE of the proposed structured
FIM-BTD estimator decrease as the SNR increases, while the baselines
exhibit larger errors, particularly at moderate and high SNR. Each
baseline reflects a distinct modeling limitation: the rigid-panel LS
estimator lacks out-of-plane geometric diversity, the static-UPA LS
estimator ignores morphing-dependent phases, and the common-steering
model cannot represent user-specific spatial responses. Calibrated
out-of-plane morphing adds sensitivity to $\sin\theta\sin\phi$, which
complements the $\sin\theta\cos\phi$ and $\cos\theta$ components
observed in the undeformed geometry. The rigid-panel comparison uses
the same total pilot-symbol overhead.

\subsection{Impact of Morphing Diversity}
\label{subsec:sim_M}

Fig.~\ref{fig:btd_M} examines how the number of morphing
configurations affects recovery of the user-specific UPA steering
vectors. The proposed structured FIM-BTD estimator achieves the lowest
NMSE over the considered range, with accuracy improving as $M$
increases. Additional configurations provide both more observations
and greater geometric diversity for estimating the coupled spatial
and morphing factors. The equal-overhead rigid-panel LS baseline also
benefits from the correspondingly longer pilot duration,
$T_{\rm rigid}=MT$, but all its observations are collected under one
fixed geometry and thus lack the directional information added by FIM
morphing. The static-UPA LS and common-steering baselines remain
limited by their respective model mismatches. Since the total number
of pilot symbols also increases with $M$, the observed improvement
reflects both additional training observations and enhanced morphing
diversity.

\subsection{Robustness to Pilot Correlation}
\label{subsec:sim_pilot_corr}
For this experiment, two users are located at
$(\theta_1,\phi_1)=(35^\circ,15^\circ)$ and
$(\theta_2,\phi_2)=(35^\circ,35^\circ)$, and the normalized pilot
correlation is defined as follows~\cite{4786505}:
\begin{equation}
\rho_s
=
\frac{
	|\mathbf s_1^H\mathbf s_2|
}{
	\|\mathbf s_1\|_2\|\mathbf s_2\|_2
}.
\label{eq:pilot_correlation}
\end{equation}

\begin{figure*}[t]
	\centering
	\subfloat[Joint angular RMSE.]{
		\includegraphics[width=0.48\textwidth]
		{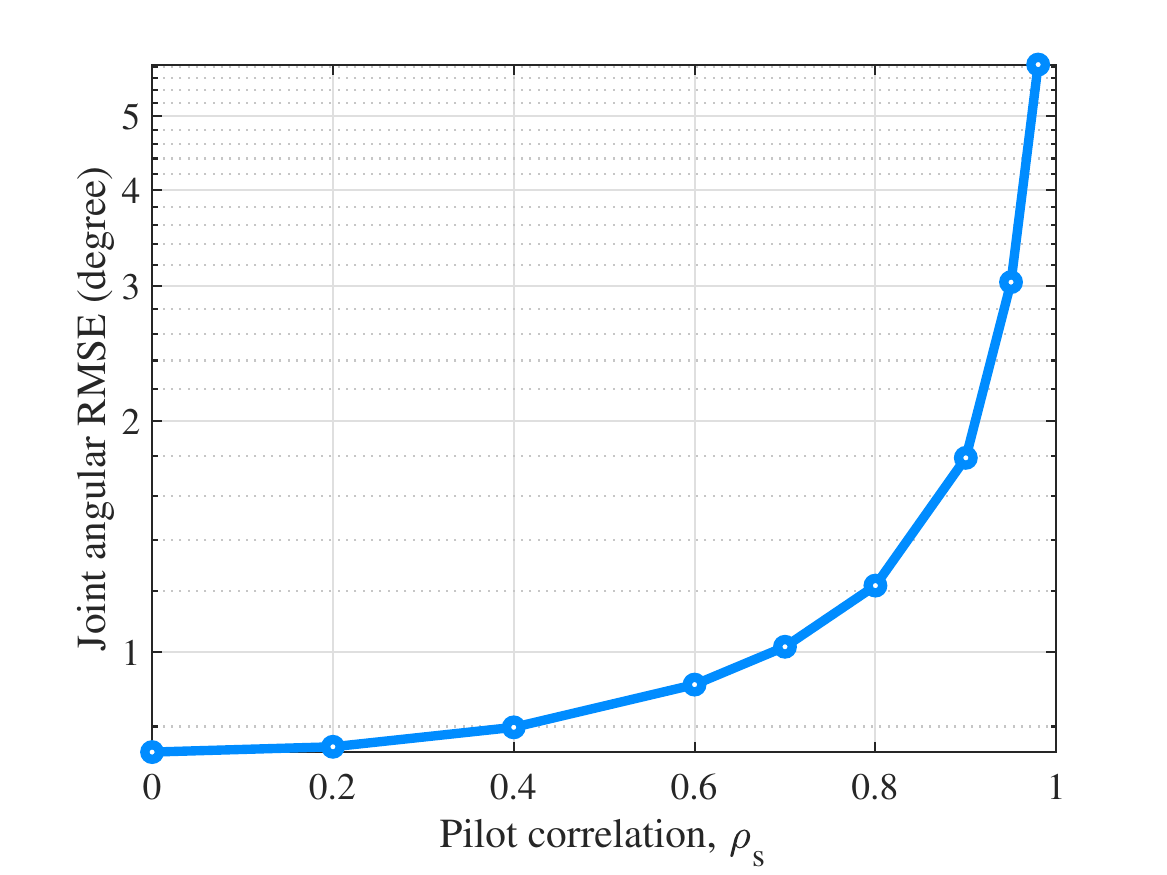}
		\label{fig:btd_rho_angle}
	}
	\hfill
	\subfloat[Final reconstructed-channel NMSE.]{
		\includegraphics[width=0.48\textwidth]
		{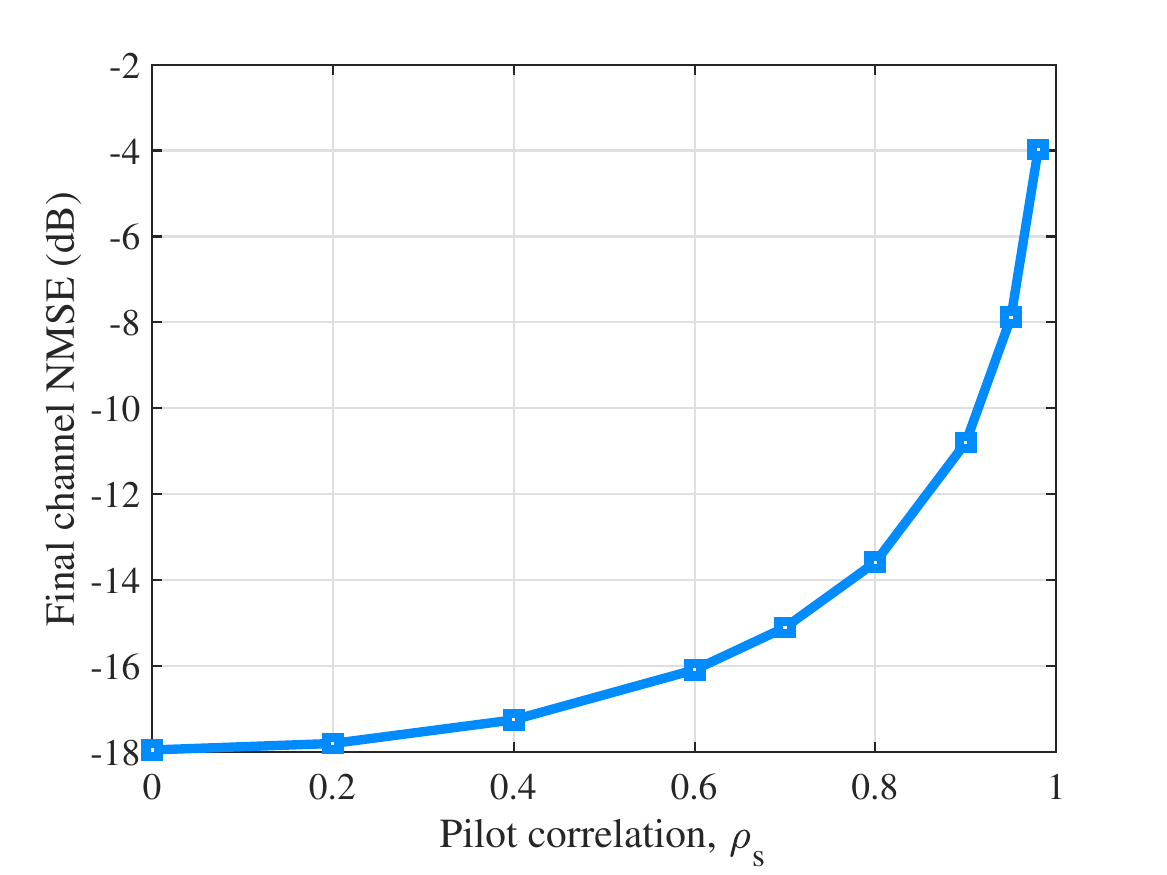}
		\label{fig:btd_rho_channel}
	}
	\caption{Robustness of the proposed structured FIM-BTD estimator
		to pilot correlation $\rho_s$: (a) joint angular RMSE;
		(b) final reconstructed-channel NMSE.}
	\label{fig:btd_rho}
\end{figure*}

Fig.~\ref{fig:btd_rho} shows the effect of pilot correlation on the
proposed structured FIM-BTD estimator. It can be seen that both the angular RMSE and
reconstructed-channel NMSE increase gradually with $\rho_s$ and
deteriorate rapidly when the pilots become highly correlated. For the
adopted two-user pilot construction with equal pilot energies, the
pilot Gram matrix has the spectral condition number
\begin{equation}
\operatorname{cond}_2(\mathbf S\mathbf S^H)
=
\frac{1+\rho_s}{1-\rho_s},
\label{eq:pilot_condition_number}
\end{equation}
As $\rho_s$ approaches unity, the condition number in
\eqref{eq:pilot_condition_number} grows rapidly. Although $\mathbf S$
remains full row rank for $\rho_s<1$, the resulting least-squares
problems become ill-conditioned and amplify noise. Thus, full row rank
is an algebraic pilot condition, not a guarantee of numerical
stability; well-conditioned pilots are also needed.

%\subsection{Impact of Multiuser Loading}
\label{subsec:sim_K}

\begin{figure}[t]
	\centering
	\includegraphics[width=\columnwidth]
	{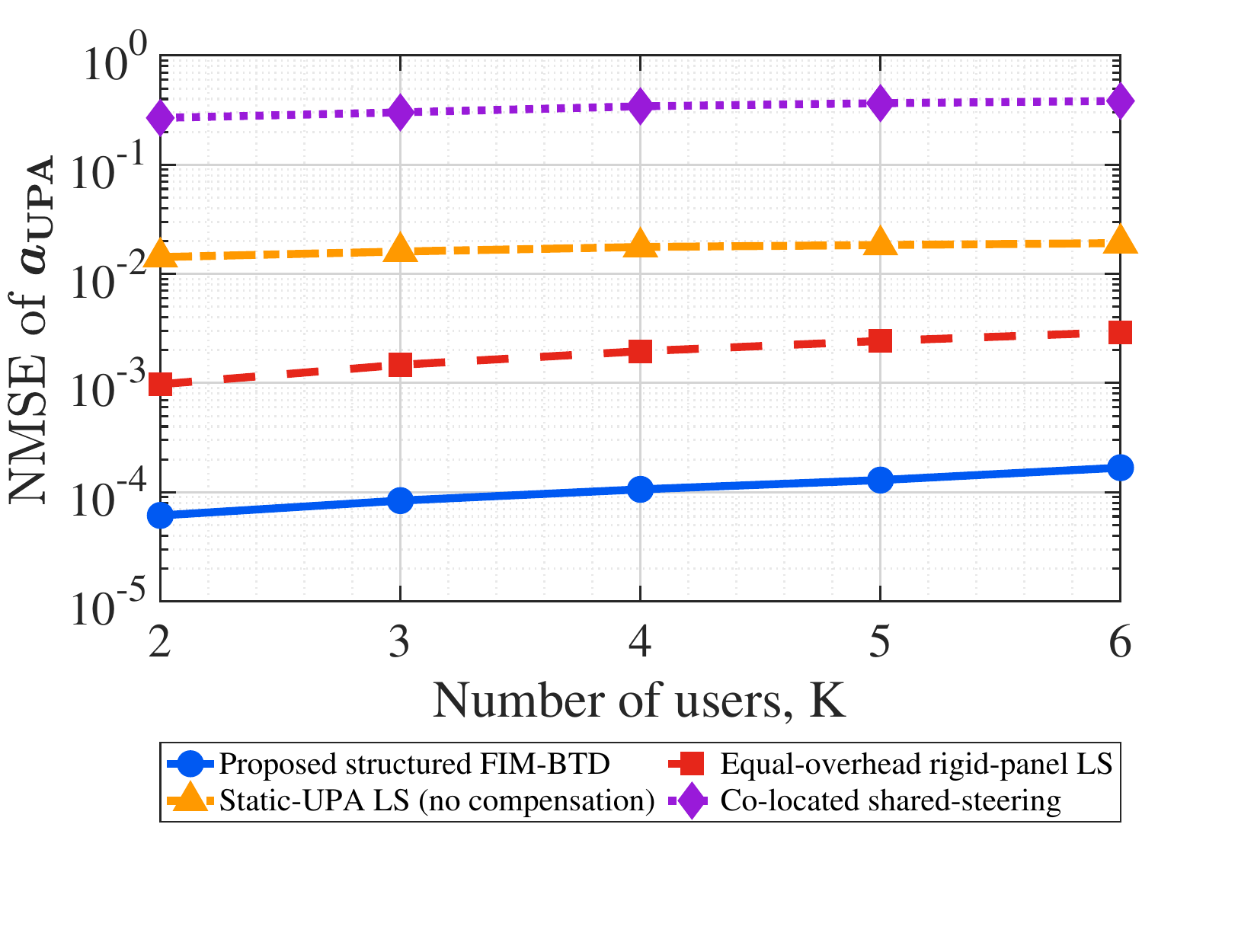}
	\caption{NMSE of the estimated user-specific UPA steering vectors
		versus the number of users $K$ with $N=16$, $M=16$, $T=12$,
		and $\mathrm{SNR}=10$ dB. The total user-gain power is normalized
		as $\sum_{k=1}^{K}|\alpha_k|^2=1$.}
	\label{fig:btd_K}
\end{figure}

Fig.~\ref{fig:btd_K} evaluates how multiuser loading affects recovery
of the user-specific UPA steering vectors. It can be noted that the estimation error
generally increases with $K$, i.e.,  more user-specific spatial--morphing
components must be recovered from the same pilot and morphing
dimensions, while the total-power normalization reduces the average
gain power per user. Nevertheless, the proposed structured FIM-BTD
estimator maintains the lowest NMSE over the considered range, with
moderate degradation as the number of users increases. This result
corroborates the benefit of jointly exploiting the known pilot
directions, diagonal spatial structure, and user-specific morphing
responses.

%\subsection{Computational Runtime}
\label{subsec:sim_runtime}

\begin{figure}[t]
	\centering
	\includegraphics[width=\columnwidth]
	{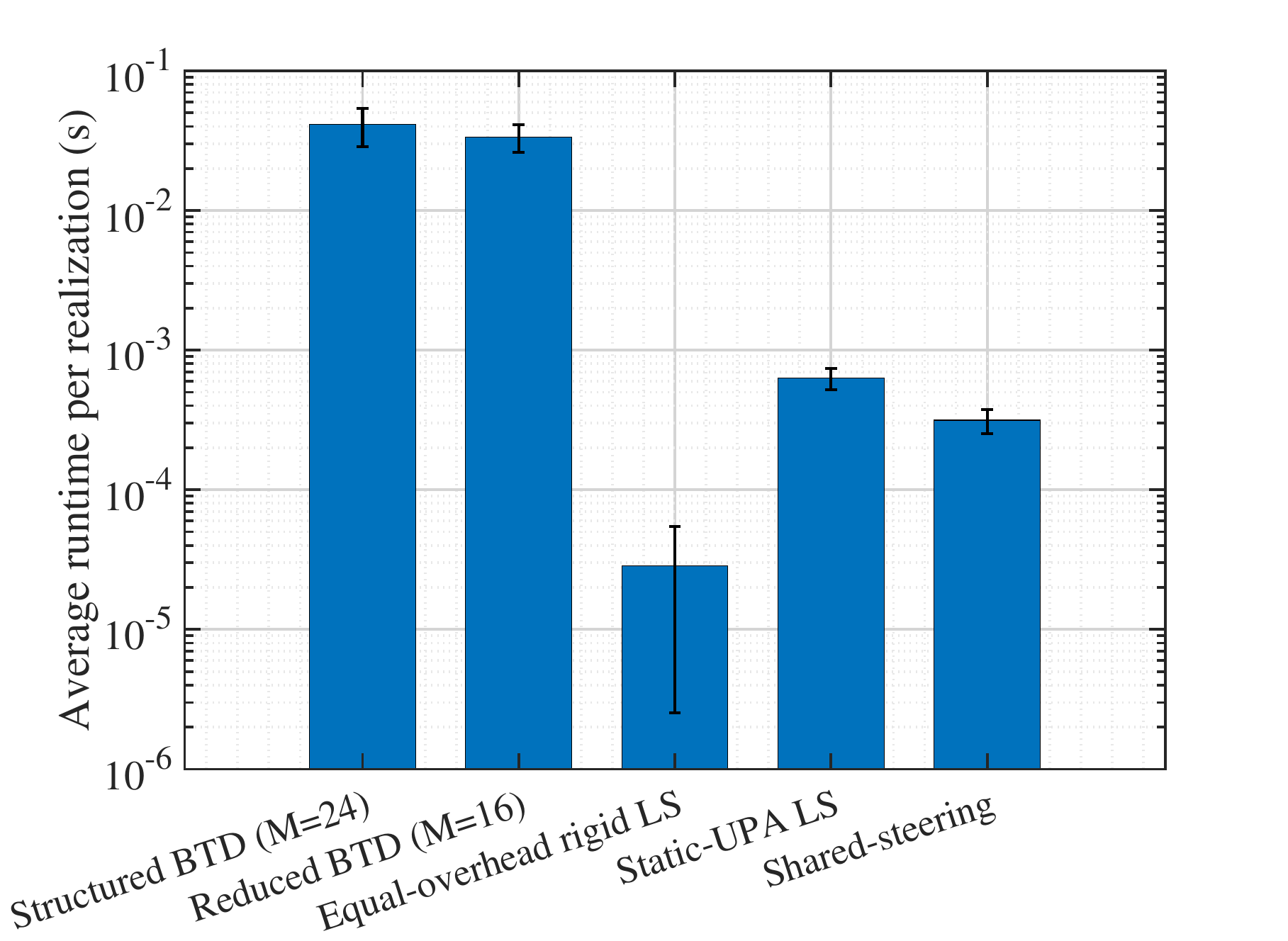}
	\caption{Average computational runtime of the considered estimators
		for the angularly separated-user scenario.}
	\label{fig:btd_runtime}
\end{figure}

Fig.~\ref{fig:btd_runtime} compares the runtimes of the considered
estimators. The structured FIM-BTD estimator requires more processing
than the LS-based baselines since it iteratively updates the
user-specific morphing factors, diagonal spatial factors, and complex
gains, followed by physical-manifold matching. Runtime decreases as
the number of morphing configurations is reduced because fewer
observations are used in the factor updates. By contrast, the
rigid-panel and static-UPA LS estimators remain computationally
inexpensive because they rely on noniterative matrix operations. These
trends agree with the complexity analysis in
Section~\ref{subsec:btd_complexity} and illustrate the tradeoff
between processing cost and the accuracy gains from exploiting the
spatial--morphing structure.
}

\section{Conclusion}
\label{sec:conclusion}

{\color{blue}
This paper develops a pilot-constrained structured BTD framework for
uplink channel estimation in sensing-enabled FIM systems with
angularly separated users. The multi-configuration observations are
modeled as a sum of user-specific blocks coupling diagonal
reference-array factors, morphing responses, and known pilot
directions. The proposed estimator alternates structured updates of the
morphing factors, spatial responses, and user gains. It uses the undeformed configuration to set the factor-scaling convention, then
matches the recovered factors to the calibrated spatial--morphing
manifold to estimate angles and reconstruct configuration-dependent
channels. We characterize rank conditions for uniqueness of each
conditional least-squares update and distinguish them from global
factor uniqueness, physical-angle identifiability, and convergence of
the nonconvex algorithm. Simulations demonstrate channel and
two-dimensional angle recovery in the tested settings, improved
steering-vector estimation as morphing diversity increases, and
degradation with increased user loading or pilot correlation. Future
work will consider multipath and wideband channels, imperfect
morphing calibration, and compressed or partially observed
element-domain measurements.
}

\bibliographystyle{IEEEtran}
\bibliography{ref}

\end{document}